\documentclass[12pt]{iopart}
\usepackage{epsf}
\usepackage{fleqn}
\usepackage{graphicx}

\newcommand{\nn}{\nonumber}
\newcommand{\be}{\begin{equation}}
\newcommand{\ee}{\end{equation}}
\newcommand{\bea}{\begin{eqnarray}}
\newcommand{\eea}{\end{eqnarray}}
\newcommand{\XI}{\mbox{\boldmath $\xi$}}
\newcommand{\PSI}{\mbox{\boldmath $\psi$}}

\begin{document}

\title{Derivation of Hebb's rule}
\author{M Heerema \footnote[6]{E-mail address: heerema@phys.uva.nl} and WA
    van Leeuwen \footnote[7]{E-mail address: leeuwen@phys.uva.nl} }
\address{Institute for Theoretical Physics, University of Amsterdam,
  Valckenierstraat 65, 1018 XE Amsterdam, The Netherlands}

\abstract
On the basis of the general form for the energy needed
to adapt the connection strengths $w_{ij}$ of a network in which learning takes place, a local learning rule is found for the changes $\Delta
w_{ij}$. 
This biologically realizable learning rule turns out to comply with
Hebb's neuro-physiological postulate, but is not of the form of any of the learning rules proposed in the literature.

The learning rule possesses the property that the energy needed in each
learning step is minimal, and is, as such, evolutionary attractive.
Moreover, the pre- and post-synaptic neurons are found to influence 
the synaptic changes differently, resulting in a asymmetric connection
matrix $w_{ij}$, a fact which is in agreement with biological observation. 

It is shown that, if a finite set of the same patterns is presented
over and over again to the network, the weights of the synapses
converge to finite values. 

Furthermore, it is proved that the final values found in this
biologically realizable limit are the same as those found
via a mathematical approach to the problem of finding the weights of a
partially connected neural network that can store a collection of
patterns.
The mathematical solution is obtained via a modified version of the
so-called method of the pseudo-inverse, and has the inverse of a
reduced correlation matrix, rather than the usual correlation matrix,
as its basic ingredient. 
Thus, a biological network might realize the final results of the
mathematician by the energetically economic rule for the adaption of
the synapses found in this article.

\endabstract


\section{Introduction}
\label{introduction}

In this article we consider some theoretical aspects of the
changes of the connections as they could take place between the nerve
cells, or neurons, of the brain.
In a learning process, these connections change continuously, and are adapted in
such a way that a particular task, e.g., the storage of patterns, is
achieved.
The answer to the question in which way the connections between
neurons actually change, in response to external stimuli, can only be
given by experiment, not via any theoretical discussion.
Although there is a lot of experimental activity related to the study
of functioning of neurons, there is not yet a unique answer to this
question: see e.g., the $1998$ review articles of Buonomano and Merzenich
\cite{buono} or Marder \cite{marder}, or the $1990$ review article of
Brown et al. \cite{brown}.

In the forties, the Canadian psychologist Hebb conjectured in his now
famous book \emph{The organization of behavior---A neuro-physiological
  Theory} \cite{hebb} that the changes of the connections between the
neurons take place according to a `neuro-physiological postulate' that
nowadays is referred to as Hebb's rule: `When an axon of cell $A$ is
near enough to excite a cell $B$ and repeatedly or persistently takes
part in firing it, some growth process or metabolic change takes place
in one or both cells so that $A$'s efficiency, as one of the cells
firing $B$, is increased'. 
Thus Hebb's rule is a quantitative statement on the enhancement of
synaptic efficiency of signal transmission, but does not state
qualitatively, by some mathematical formula, to what extend.   

Nowadays, there is a great amount of evidence that synapses do indeed
change in a learning process, and, since the appearance of Hebb's
article many quantitative  proposals, all complying with  Hebb's
postulate, have been put forward. 
Also the present article is concerned with such a quantitative
expression for the synaptic changes.
However, rather than postulating a learning rule, we derive it from some
underlying principle. 
As a final result, we find a learning rule for the
adaptation of the strengths, or weights, $w_{ij}$, of a synapse
connecting a post-synaptic neuron $i$ and a pre-synaptic neuron $j$. 
Its explicit form reads:
\be
\label{main-result}
\Delta w_{ij}(t_n)=\eta_i [\kappa-\{h_i(t_n)-\theta_i\}(2\xi_i-1)]
(2\xi_i-1)\xi_j  
\ee
This ---asymmetric--- learning rule gives $\Delta w_{ij}$, the positive or
negative increment of the weight $w_{ij}$, as a function of the
activities $\xi_i$ and $\xi_j$  of neurons $i$ and $j$ of the synapse
that  connects these neurons. 
In our convention, the activity $\xi$ of a neuron equals $1$ if it
generates an action potential, and $0$ if it is quiescent. 
The function $h$ is the potential difference
between the interior and the exterior of a neuron, at its axon
hillock. 
The formula gives the change at time $t_n$. 
The index $n$ denotes the time at the $n$-th learning step in the
process of learning  ($n=1, 2, \ldots$). 
The threshold potential, $\theta_i$, is a constant, typical for the
neuron $i$ in question. 
It equals, by definition, the potential that must be surmounted, at
the axon hillock of neuron $i$, in order that it will fire. 
The quantities $\eta_i$ and $\kappa$ are also constants. 
Their precise identification, as variables related to individual and
collective neuron properties, is outside the scope of the present
article. 
The learning rule (\ref{main-result}), which constitutes our main result as far
as biology is concerned, has a form that is compatible with Hebb's
postulate. 

It is a well-known fact that, for a given  neural net with  strengths
$w_{ij}$ of the weights, there are \emph{infinitely} many ways to
choose changes $\Delta w_{ij}$  of the weights such  that the network
will perform storage and retrieval of a new pattern. 
The derivation of our learning rule is based on the assumption that,
at each instant of the learning process, the energy needed  to change
the neural network in order to store a new pattern, is minimal.  
The requirement that, at each step $n$ of the learning process, the
energy energy needed is as low as possible, turns out to be sufficient to
\emph{uniquely} determine the way in which the weight of each synapse
connecting two arbitrary neurons $i$ and $j$ should be changed, and
thus fixes a learning rule for the adaptation of the weights of all the
connections. 
We will call this learning rule, the `non-local energy saving learning rule', since it
turns out to depend on the state of activity of \emph{all} neurons
$j$ from which neuron $i$ receives its input. 
It is given by eq. (\ref{delta-lagrange}) below. 

It is impossible, however, for a synapse connecting two neurons $i$
and $j$, to realize the non-local energy saving learning rule
(\ref{delta-lagrange}) exactly, as follows by a careful inspection of
formula (\ref{delta-lagrange}). 
In fact, in order to adapt itself according to this learning rule, a synapse
between $i$ and $j$  would have to `know' the individual states of
activity $\xi_k$ of \emph{all} pre-synaptic neurons $k$ from which
neuron $i$ gets its input, whereas a synapse only `feels' the states of
the two neurons $i$ and $j$ which it connects. 
The best a synapse can do in order to compete with  the performance of
the non-local learning rule (\ref{delta-lagrange}) is to adapt itself
according to a learning rule that is a local approximation of the non-local learning rule. 
It is this local approximation, given the expression
(\ref{main-result}) above, which constitutes our  main biological
result. 
We will refer to it as the local energy saving learning rule, to distinguish
it from its non-local counterpart. The point of locality of learning
rules is discussed in more detail in section~\ref{local}. 

A numerical estimation of the performance of the local learning rule,
eq. (\ref{main-result}), versus to the non-local one,
eq. (\ref{delta-lagrange}), is made in section~\ref{numerical}. 
Local learning turns out to be a very effective alternative for
non-local learning, as well as regarding its power to store and
retrieve patterns as with respect to its capacity to be economic in
use of energy. 

In order to arrive at the non-local energy saving learning rule, we think of a
neuron as a living cell. A living cell, as a physical object, is a
stationary non-equilibrium system. 
The basic assumption of this article is that any type of change of the
cellular cleft can only be effected by \emph{adding} energy to the
non-equilibrium system, independent of the fact whether it leads to a
strengthening or a weakening  of the synaptic efficacy. 
This is a plausible, but not totally trivial postulate, which can only
be falsified by a detailed biophysical or biochemical study of the
process of change of the synapse. 
In our setup, the mere assumption that extra energy is needed for any
change of the synapse, independent of the fact whether it leads to an
increase or a decrease of its efficiency, replaces Hebb's postulate on
efficiency cited above. 

Before starting the derivation of the energy saving learning rule itself, we
discuss, in section~\ref{hebb}, the $81$ possibilities which, in
principle, are compatible with Hebb's postulate. 
In particular, we consider these mathematical realizations with respect to there
biological plausibility. 
We then find that, in fact, out of the $81$ learning rules that are
possible in principle, only two are also biologically plausible. 
These are the learning rules (\ref{the-one}) and (\ref{other}). 

The actual derivation of the energy saving learning rule is performed in
section~\ref{lagrange}. 
To our satisfaction, it general form turns out to imply the two forms
(\ref{the-one}) and (\ref{other}) expected in the preceeding section
on biological grounds only. 
Thus our `principle of minimal change of energy', which might lead, a
priori, to any of the $81$ possibilities for a realization of a learning rule
for the change of weight of a synapse, happens to yield precisely
those rules which are biological plausible.  

In section~\ref{ideal} we consider the situation that the changes of the
connections do not take place in an energetically optimal way, but in
such a way that patterns are not partially wiped out when new patterns
are learned as is the case for learning based on the energy saving
learning rule (\ref{main-result}) or (\ref{delta-lagrange}). 
We then ask ourselves the question which learning rule would then be found for
the changes $\Delta w_{ij}$ of the synaptic weights. 
Again, its general form turns out to comply with one of the $81$
possible realizations of the Hebb rule considered in section~\ref{hebb}, but,
in this case, it is an biologically improbable one. 
We therefore do not pursue this path any further.

The question might arise whether the non-local energy saving learning rule
converges, in the limit that the number of learning steps tends to
infinity. 
And, if so, to what values they then would converge. 
The answers to these questions are the subject of section~\ref{p>1}. 

There exists a well-know way to obtain the final form of the
connection strengths $w_{ij}$ of an artificial neural network
that can store and retrieve a set of patterns: it goes under the name
`pseudo inverse solution' \cite{personnaz,kohonen}. 
By inversion of a certain matrix related to the patterns to be stored,
the  so-called correlation matrix, one can obtain, without any
limiting procedure, final values for the weights $w_{ij}$ of the connections of
a neural network that yield the desired result of being capable of
storing and retrieving a collection of patterns.  

We will consider an assembly of $N$ neurons, where $N$ is a number
relevant for a certain subunit of the brain, such as a cortical
hyper-column, for which $N$ is of the order of $10^4$ to $10^5$. 
Although such subunits are highly interconnected, they are partially
connected in
the mathematical sense, since each neuron is connected to only a
finite fraction of the subunit considered. 
Moreover, biological neurons are not self-connected, i.e.,
$w_{ii}=0$. 
These two biological facts force us to study, from the very beginning,
diluted, or partially connected, networks.  
In the limit that the dilution tends to zero, we rediscover, if we
relax the requirement that  the self-connections all vanish, some of
the well-known results for fully connected networks, in particular
those of Diederich and Opper \cite{opper}, and of Linkevich \cite{linkevich}. 

A possible question one might now ask is: is there any relation
between the final values obtained for the weights $w_{ij}$ obtained in
the limit of an infinite number of learning steps, $n\rightarrow \infty$, at the one hand and the values
obtained via the pseudo-inverse method at the other hand? 
The answer to this question is as simple as it is amazing: the results
are identical.  
The proof of this point is the subject of appendix B, where the method
of pseudo-inverse is modified in such a way that it can be used for
partially connected networks.
Thus, as a final conclusion, we can state that \emph{i}. the
assumption of economy of energy in a learning step, \emph{ii}. the
well-known method based on the pseudo-inverse of the correlation
matrix and \emph{iii}. biological plausibility of a learning rule are
three members of a trio that work in concert.  
We want to stress, once again, that the question whether the
evolutionary development of  the brain actually has led to  an
adaptation process of the synapses  that is energetically the most
economical, is, as yet, experimentally, an open question. 
It is not excluded that the realization of the changes of the synapses
might take place in a biologically less probable, or an energetically
less favourable way. 
Our only certainty is  that  economy of energy and  biological
probability go hand in hand. 

Usually, neural networks have been modeled in the so-called
spin-representation, which, in principle, can easily be translated to
the so-called binary representation, which models the biological
reality more directly. 
In particular, in the binary representation the thresholds for activation of a
neuron can be taken constant, in accordance with the biological
reality.
In the spin-representation, however, the actual biological reality in
a learning process can only be
modeled via the use of a time-dependent threshold, a fact which is
often overlooked: one erroneously treats the neuron thresholds in the
spin model as constants, see, e.g., \cite{kanter,krauth}.
We therefore have chosen not to use the spin, but the binary representation.

In our study of the connections $w_{ij}$ and the way in which they
change in a learning process, we will neglect two constraints set by
nature.
Firstly, the fact that, for an actual neuron $i$, the magnitudes of
the synaptic connections are within some interval characteristic for
the synapse in question.
Secondly, the fact that, according to Dale's law, the connections
related to one and the same pre-synaptic neuron either are only
excitatory or only inhibitory.
Furthermore, we treat biological neurons as McCulloch and Pitts
neurons, i.e., their response to input is according to the rule
(\ref{neuron-model})--(\ref{dynamical-rule}) below.
We thus also neglect the retardation which results from the finite
speed of transmission of signals through axons and dendrites.
A way retardation could be included in a model has been put forward in
\cite{hemmen-ritz}.

For an introduction to this article, see textbooks such as
\cite{muller,domany,hertz}.

\section{Attractor neural network model}
\label{network}

\paragraph{Dynamics}

We consider a network of $N$ interconnected neurons in the binary
representation, i.e., each neuron can have a state $x_i=1$ (the neuron
produces \emph{one} action-potential or \emph{spike}) or $x_i=0$ (the neuron is quiescent).
The post-synaptic potential of neuron $i$ at time $t$ of this
system of neurons is modeled by 
\be
\label{neuron-model}
h_i(t)=\sum_{j=1}^N w_{ij}(t)x_j(t) \, , \quad (i=1,\ldots,N) \, ,
\ee
where the $x_j(t)$ are the input signals at time $t$ and where the
$w_{ij}(t)$ are the \emph{weights}, also called \emph{synaptic
  strengths} or \emph{synaptic efficacies} at time $t$.
A weight $w_{ij}$ takes into account the overall effect of a
synaptic connection between a post-synaptic neuron $i$ and a
pre-synaptic neuron $j$ and may be positive (excitation),
negative (inhibition) or zero (no synaptic connection).
The weights $w_{ij}$, like the potentials $h_i$, are expressed in
Volts.
The output of neuron $i$ is supposed to be given by the dynamical equation
\be
\label{dynamical-rule}
x_i(t+\Delta t)={\rm \theta}_H \{ h_i(t)-\theta_i\} \, , \quad
(i=1,\ldots,N) \, ,
\ee
where the \emph{constant} $ \theta_i$ is the activation threshold characteristic of
neuron $i$ and where $\Delta t$ is some discrete time step.
A typical value for $\theta_i$ is $10$ ${\rm mV}$ \cite{kandel}.
The symbol $\theta_H$ stands for the Heaviside step function, which
equals one for positive arguments and vanishes otherwise.

In the so-called `spin-representation', active and non-active states
of neuron $i$ are characterized by $s_i=1$ or $s_i=-1$, respectively.
In this representation, the dynamical equation (\ref{dynamical-rule})
can be rewritten as
\be
\label{dynamical-rule-spin}
s_i(t+\Delta t)={\rm sgn} \{ \sum_{j=1}^N J_{ij}(t)s_j(t) -T_i(t) \} \, , \quad
(i=1,\ldots,N) \, ,
\ee
where the time dependent `coupling constants' $J_{ij}$ are related to the biological
weights $w_{ij}$ through $J_{ij}=w_{ij}/2$ and where $s_j=2x_j-1$.
The time dependent `thresholds' $T_i(t)$ are related to the constant
biological thresholds $\theta_i$ according to
\be
\label{def-t}
T_i(t)=\theta_i - \sum_{j=1}^N J_{ij}(t) \, , \quad (i=1,\ldots,N) \, .
\ee
In the literature the thresholds $T_i(t)$ are usually treated as a
constant; most often the constant is taken to vanish \cite{kanter,krauth}.
This seemingly innocent fact changes, of course, the dynamics
(\ref{dynamical-rule-spin}) of the system in a non-trivial way.
As a consequence, the results obtained for, e.g., the adaptation of
the coupling constants differ from those obtained when the
actual biological dynamics (\ref{dynamical-rule}) is used
[cf. eqs. (\ref{spin-binary}) and (\ref{spin-spin})].
Hence, when modeling adaptation processes of biological neurons with
constant thresholds, the use of the binary representation is obligatory.

Neural networks have two time scales, one related to the rate of
change of the synaptic efficacies $w_{ij}$ and one related to the
spiking activity of a neuron.
The latter time is of the order of milliseconds, the former is less
well defined, but can be estimated to lie somewhere between seconds
and days: it is a time related to the rate of learning of a brain.
Hence, the $\Delta t$ occurring in equation (\ref{dynamical-rule}) is
of the order of milliseconds.
When the process of adaptation of the weights has come to an end the
$w_{ij}$ remain constant.

\paragraph{Fixed points}

We want to determine the synaptic efficacies of an attractor neural
network, i.e., of a network which can recall a number, $p$ say, of
previously stored patterns.
The realization of a recall corresponds to a fixed network state of
the network dynamics (\ref{dynamical-rule}).
Let us denote the patterns of activity, or patterns, by
$\XI^{\mu}=(\xi_1^{\mu},\ldots,\xi_N^{\mu})$, where $\mu=1,\ldots,p$.
Thus $\xi_i^{\mu}=1$ or $\xi_i^{\mu}=0$ with $i=1,\ldots,N$ and
$\mu=1,\ldots,p$.
The probability that a neuron $i$ is in the state $1$
or $0$ is supposed to be given by $a$ or $(1-a)$ respectively.
The quantity $a$ is usually called the mean activity of the neural
net.
For random patterns the mean activity $a$ is given by $0.5$.
In biological neural networks, however, the mean activity $a$ is
smaller \cite{abelles}.

Thus, a network which has stored, somehow, $p$ patterns $\XI^{\mu}$
satisfies the fixed point equations
\be
\label{def-fixed-point}
x_i (t+\Delta t)=x_i(t) \, , \quad {\rm for} \quad x_i(t)=\xi_i^{\mu} \, , \quad (i=1,\ldots,N;\mu=1,\ldots,p) \, ,
\ee
Hence, equations (\ref{dynamical-rule}) and (\ref{def-fixed-point}) yield
the $pN$ equations
\be 
\label{fp}
\xi_i^{\mu}={\rm \theta_H}\{\sum_{j=1}^N w_{ij}(t)\xi_j^{\mu}-\theta_i
\}
\ee 
for $N^2$ unknown $w_{ij}$'s.

Let us now introduce so-called stability coefficients $\gamma_i^{\mu}$
\cite{kinzel}:
\be
\label{gamma}
\gamma_i^{\mu}(t):=\left( h_i^{\mu}(t)- \theta_i \right)
\left( 2\xi_i^{\mu}-1 \right) \, ,
\ee
with $h_i^{\mu}$ the post-synaptic potential
\be
\label{h-mu}
h_i^{\mu}(t)=\sum_{j=1}^N w_{ij}(t)\xi_j^{\mu} \, .
\ee
Remark that $\gamma_i^{\mu}$ depends, via $h_i^{\mu}$, on all
weights $w_{ij}$, i.e., $ \gamma_i^{\mu}(t)= \gamma_i^{\mu}(w_{11}(t),
w_{12}(t),\ldots,w_{N-1,N}(t),w_{NN}(t)) $.

One easily checks, by distinguishing the cases $\xi_i^{\mu}=1$ and
$\xi_i^{\mu}=0$, that an equivalent way to express the equalities
(\ref{fp}) are the $pN$ \emph{in}equalities 
\be
\label{fp-equiv}
\gamma_i^{\mu}(t) > 0 \, .
\ee
The inequality sign in (\ref{fp-equiv}) reflects that fact that the
set of equations
(\ref{fp}) is under-determined, i.e., the eqs. (\ref{fp-equiv}) are necessary but not sufficient equations to determine
uniquely a set of weights of a network which has stored some patterns.

An arbitrary pattern ${\bf X(t)}$ will only be recalled if it evolves
in time to one of the fixed points $\XI^{\mu}$.
Therefore, it is not sufficient for a
network to have fixed points: for each of the $p$ fixed points that is
related to a retrieval of a pattern $\XI^{\mu}$, there must exist a
whole neighborhood of points around $\XI^{\mu}$ which is such that all
points of this neighborhood will evolve to $\XI^{\mu}$ under the dynamics
(\ref{dynamical-rule}).
In technical terms, the fixed points $\XI^{\mu}$ must have a non-zero
basin of attraction.
For this reason, one may introduce \cite{opper,krauth,gardner} a positive
threshold $\kappa$, and demand the stronger inequalities 
\be
\label{stable}
\gamma_i^{\mu}(t) \ge \kappa
\ee
to hold, rather than the inequalities (\ref{fp-equiv}), which are
equivalent to the fixed point equations (\ref{fp}).
The larger the threshold $\kappa$, the larger the basins of attractions can
be expected to be \cite{krauth,gardner}.

In order to solve the equations (\ref{stable}) for the unknown weights $w_{ij}$, we consider it as far as its equality sign is concerned.
Then (\ref{stable}) can be recast in the equivalent form
\be
\label{linear-fp}
\sum_{j=1}^N w_{ij}(t)\xi_j^{\mu}-\theta_i=\kappa(2\xi_i^{\mu}-1) \, ,
\quad (i=1,\ldots,N;\mu=1,\ldots,p) \, ,
\ee
as may be checked by putting $\xi_i^{\mu}$ equal to $1$ or $0$.
The $pN$ equations (\ref{linear-fp}) do not fix uniquely the $N^2$
weights $w_{ij}$ as long as $p<N$, the case we consider throughout
this article.
The storage capacity $\alpha$, defined as $\alpha:=p/N$, of a neural
network is maximally equal to one for networks described by
eqs. (\ref{linear-fp}).

\paragraph{Various types of networks}
\label{types}

It is our aim to take into account specific aspects of the connectivity of a biological network.
In a biological neural network a neuron does not excite or
inhibit itself, i.e., for all $t$ we have for the self-interactions
(or self-connections) 
\be
w_{ii}(t)=0 \, ,\quad (i=1,\ldots,N) \, .
\ee
Moreover, a biological network will, in general, be partially connected: each neuron will have some neighbourhood outside which
there are no connections, i.e.,
\be
w_{ij}(t)=0 \, ,
\ee
for a given set of neuron pairs ($i,j$).
We shall call a network in which a (finite)
fraction of the weights vanish, a diluted network.
Let $M_0$ be the number of pairs ($i,j$) for which $w_{ij}(t)=0$.
Then the dilution $d$ of a network of $N$ neurons is defined as
\be
\label{d}
d:= M_0/N^2 \, .
\ee
Hence, the dilution $d$ is a number between $0$ and $1$.

Let us slightly generalize the above by distinguishing in a learning
process changing and non-changing connections $w_{ij}(t)$ instead of changing and
vanishing connections.
Let us consider, for a moment, one particular neuron $i$.
Then one may define the index sets
\be
\label{v}
V_i:=\left\{ j \ | \ w_{ij}(t) \neq w_{ij}(t_0) \right\} \, , \qquad
V_i^c:=\left\{ j \ | \ w_{ij}(t)=w_{ij}(t_0) \right\} \, .
\ee 
Thus $V_i$ contains the indices related to all connections of neuron
$i$ that, in a learning process, change in time, whereas its complement, $V_i^c$, contains the
indices related to all non-changing connections.
In particular $V_i^c$ contains the index of neuron $i$ itself
[$w_{ii}(t)=w_{ii}(t_0)=0$], the indices of neurons $j$ which have no
connections with neuron $i$ [$w_{ij}(t)=w_{ij}(t_0)=0$], and the
indices of neurons $j$ which have connections with fixed strengths with neuron $i$
[$w_{ij}(t)=w_{ij}(t_0) \neq 0$].
Thus, diluted networks are a subclass of networks with
changing and non-changing connections.
By specifying, via eq. (\ref{v}), which connections are absent, the
network connectivity is completely defined.
For later use, we introduce $M$, the number of pairs ($i,j$)
for which $w_{ij}(t)=w_{ij}(t_0)$ is constant, but not necessarily
equal to zero.

\section{Learning prescriptions --- Hebb rules}
\label{hebb}

In this section we will consider all mathematical realizations which
are, in principle, compatible with Hebb's postulate.
We will argue that, in our view, only two of them, namely
(\ref{the-one}) and (\ref{other}) are biologically plausible, in
contrast to the realizations (\ref{asym}) and (\ref{sym}) used in the
literature.
In order to show this, let us consider a network with changing and non-changing connections,
in which a learning process takes place with the purpose to store a collection of $p$ patterns $\XI^{\mu}$.
Let the weights at time $t_n$ be given by $w_{ij}(t_n)$.
After a learning step the new weights will be given in terms of the
old weights by
\be
\label{learning-step}
w_{ij}(t_{n+1})= \left\{ \begin{array}{ll} w_{ij}(t_n)+\Delta
w_{ij}(t_n) \, ,& (j \in V_i) \\ w_{ij}(t_n) \, ,& (j \in V_i^c)
\end{array} \right.
\ee 
where $\Delta w_{ij}(t_n)$ is the increment at time $t_n$.
A learning rule is a recipe for the change $\Delta w_{ij}$ as a
function of the states of the post-synaptic neuron $i$ and the
pre-synaptic neuron $j$ when a pattern
($\xi_1,\ldots,\xi_N$) is presented to the network.
There are four possible states ($\xi_i,\xi_j$) that the
post- and pre-synaptic neuron can have, namely
($0,0$), ($0,1$), ($1,0$) and ($1,1$), each of which may lead
to one of the three possible changes for $\Delta w_{ij}$: positive,
negative or zero.
Hence, in principle there are $3^4=81$ possible learning rules
\be
\label{mapping}
\Delta w_{ij}: (\xi_i,\xi_j) \longmapsto \Delta
w_{ij}(\xi_i,\xi_j) \, .
\ee
\begin{table}[htb]
\epsfxsize=15cm
\centerline{\hbox{\epsffile{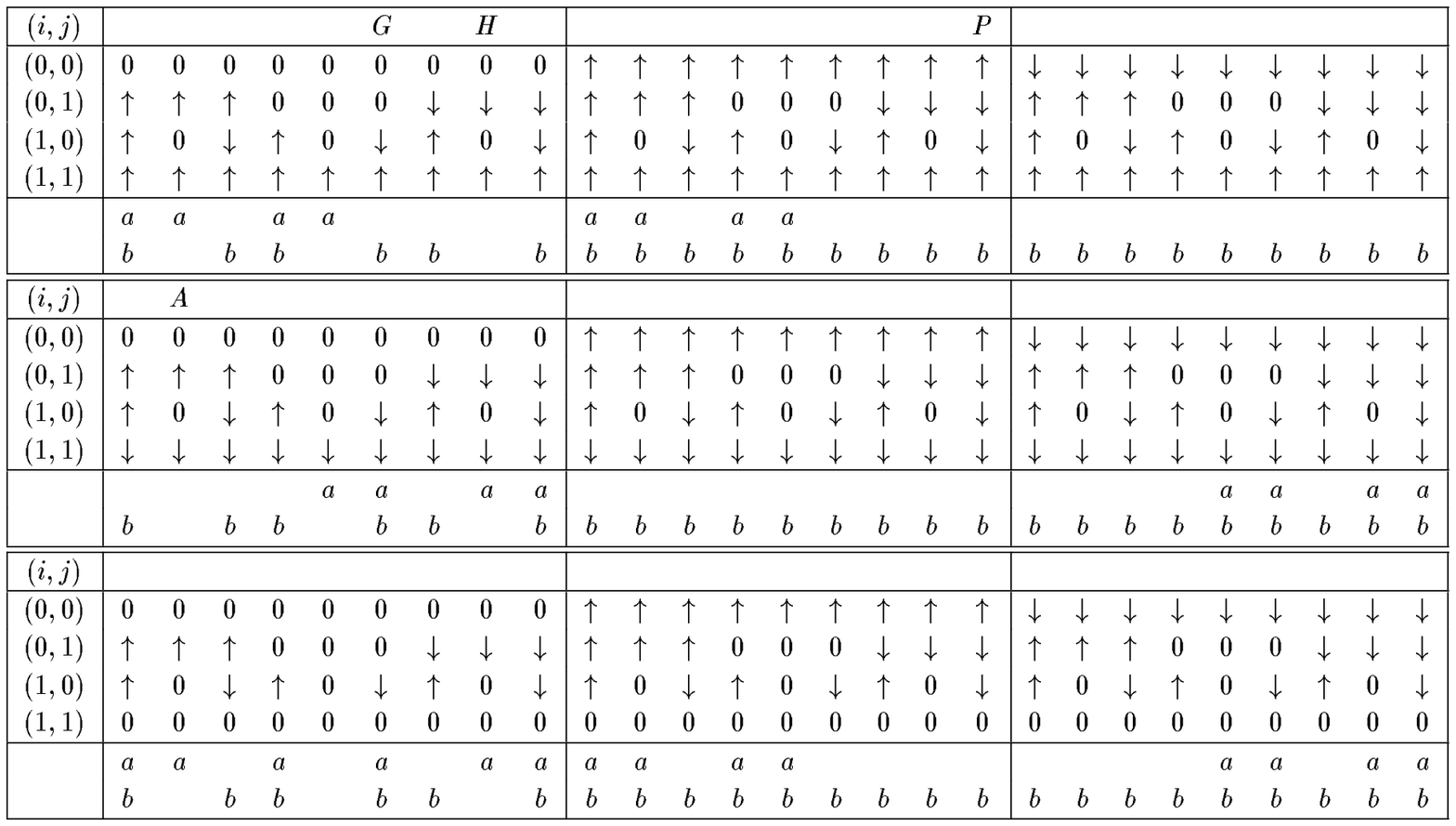} }}
\protect\caption[0]{The $81$ possible ways in which $w_{ij}$ may change as
  a function of the activities of the post-synaptic neuron $i$ and the
  pre-synaptic neuron $j$ can be read off from the $81$ columns of the
  table. Each row may have up arrows ($\uparrow$), down arrows
  ($\downarrow$) or zeros, indicating a strengthening, a weakening or
  no change of a synaptic connection. The biological reason to reject a column is
  indicated by the letter \emph{a} or \emph{b} immediately below the column. The reasons are \emph{a}: there either is only strengthening or
  weakening of the synapse, \emph{b}: there is a change of the synaptic
  strength if the pre-synaptic neuron $j$ is inactive. From the table we
  can read off that $78$ possibilities are excluded for  reason \emph{a}
  and/or \emph{b}. The column with only zeros is excluded for obvious reasons. The two possibilities for the Hebb rule which we are left with are indicated by the symbols $H$ and $A$: the first  corresponds to what is called Hebbian learning, the second to what is called anti-Hebbian learning.
If we do not reject a possibility for reason  $b$, there are many more possible Hebbian rules. The possibility indicated by $G$ was used by Gardner \protect\cite{gard-mert}. The one  preferred by physicists in their modeling of neural networks,  has been indicated by the symbol $P$.
\protect\label{tabel}}
\end{table}

It is biologically improbable that connections will always
grow or will always decrease.
Therefore, we exclude learning rules for which $\Delta
w_{ij}(\xi_i,\xi_j)$, for all four states ($\xi_i$,
$\xi_j$), are either always positive, or always negative (reason
of rejection $a$ of table~\ref{tabel})~.
Moreover, in our opinion, it is biologically probable that a
connection between a pre-synaptic neuron $j$ and a post-synaptic neuron $i$ does
not change if the neuron $j$ does not contribute to the post-synaptic
potential of neuron $i$, i.e., if
$\xi_j=0$.
Therefore, we exclude learning rules for which $\Delta
w_{ij}(\xi_i,\xi_j=0) \neq 0$ with $\xi_i=0,1$ (reason of rejection $b$ of table~\ref{tabel})~.

Excluding these improbable learning rules, we are left with no more
than two learning rules, as may be verified by a simple
inspection of table~\ref{tabel}.
One of these corresponds to the assignments 
\be
\begin{array}{ll} (0,0) \longmapsto \Delta w_{ij}=0 \, , & (0,1) \longmapsto
  \Delta w_{ij}<0 \\  (1,0) \longmapsto \Delta w_{ij}=0 \, , & (1,1) \longmapsto
  \Delta w_{ij}>0 \, , \end{array}
\ee
(column $H$ in table~\ref{tabel}), which can be expressed compactly by the formula
\be
\label{the-one}
\Delta w_{ij}=\epsilon_{ij} (2\xi_i-1) \xi_j \, ,
\ee
where the $\epsilon_{ij}$, here and elsewhere in this article, are
positive numbers.
Similarly, the other one can be expressed by the formula
\be
\label{other}
\Delta w_{ij}=-\epsilon_{ij} (2\xi_i-1) \xi_j \, .
\ee
(column $A$ in table~\ref{tabel}).

Learning can be classified as Hebbian or anti-Hebbian.
Hebbian learning is characterized by the fact that, if both neurons
$i$ and $j$ are active, $\Delta w_{ij}$ is positive, whereas for
anti-Hebbian learning $\Delta w_{ij}$ is negative.
So, the two remaining learning rules (\ref{the-one}) and (\ref{other})
are Hebbian and anti-Hebbian, respectively.
The learning rules (\ref{the-one}) and (\ref{other}) have, to the best of our knowledge, not been used, as yet, in mathematical or
physical studies
that tried to model biological neural systems (see,
e.g. \cite{hemmen-ritz,amit}).

If we allow for the possibility that $\Delta w_{ij} \neq 0$ if the
pre-synaptic neuron $j$ is inactive ($\xi_j=0$), there are many
extra possible mappings (\ref{mapping}), of which we mention the two
most often encountered in the literature
\bea
\label{asym}
\Delta w_{ij} &=& \epsilon_{ij} \xi_i(2\xi_j-1) \\
\label{sym}
\Delta w_{ij} &=& \epsilon_{ij} (2\xi_i-1) (2\xi_j-1) 
\eea
The learning rule (\ref{asym}) was used, e.g., by Gardner
\cite{gard-mert} in studying the retrieval properties of a neural
network with an asymmetric learning rule (row $G$ in table~\ref{tabel}). 
The learning rule (\ref{sym}) is the one most often used by physicists
\cite{amit,hopfield} in their modeling of neural networks (row $P$ in
table~\ref{tabel}).

Finally, let us compare the four learning rules
(\ref{the-one})--(\ref{sym}) after one learning step of one pattern
$\XI$.
Let us suppose that a pattern $\XI$ is not yet learned at time
$t_0$ so that,
in view of (\ref{stable}), the quantity $\gamma_i(t_0)$ is negative.
In order to store a pattern, $\gamma_i$ should be positive.
Upon substitution of the Hebbian or symmetric learning rules
(\ref{the-one}) or (\ref{sym}) into
(\ref{gamma}) we find
\be
\label{gamma-the-one}
\gamma_i(t_1)= \gamma_i(t_0)+
\sum_{j \in V_i} \epsilon_{ij} \xi_j \, ,
\ee
for the anti-Hebbian learning rule (\ref{other}) we get
\be
\label{gamma-other}
\gamma_i(t_1)= \gamma_i(t_0)-
\sum_{j \in V_i} \epsilon_{ij} \xi_j \, ,
\ee
whereas for the asymmetric learning rule (\ref{asym}) we obtain
\be
\label{gamma-asym}
\gamma_i(t_1)= \gamma_i(t_0)+
\xi_i \sum_{j \in V_i} \epsilon_{ij} \xi_j \, ,
\ee
where $t_0$ is the initial time and $t_1$ is the time after one
learning step.
By a suitable choice for $\epsilon_{ij}$ it can \emph{always} be achieved
that $\gamma_i(t_1)$ is positive in case of the Hebbian and symmetric
learning rules (\ref{the-one}) and (\ref{sym}), whatever are the values of
$\xi_i$ and $\xi_j$, as follows from (\ref{gamma-the-one}).
This can \emph{never} be achieved in case of the anti-Hebbian learning rule
(\ref{other}), as is seen from (\ref{gamma-other}).
Finally, in case $\xi_i=0$, this can \emph{never} be achieved for
the asymmetric learning rule (\ref{asym}), as can be read off from (\ref{gamma-asym}).
These simple arguments show that the Hebbian and symmetric learning
rules (\ref{the-one}) and (\ref{sym}) --- but
not the anti-Hebbian and asymmetric learning rules (\ref{other}) and
(\ref{asym}) --- are, in principle, suitable for storage of patterns.

In the next section we will show that the requirement that synaptic
changes take place in an energetically economic way leads to a
learning rule which, depending on the state of the post-synaptic
neuron $i$, is of the Hebbian or anti-Hebbian form (\ref{the-one}) or (\ref{other}).
Hence, the naive approach of this section, which leads to the two
forms (\ref{the-one}) and (\ref{other}), is consistent with an
approach which is based on a physical principle.

\section{Energy saving learning rule}
\label{lagrange}

In the literature, Hebb rules for the change of the synaptic connections have
been derived in various manners, many of which essentially correspond
to the determination of an extremum of some `Lyapunov' or `cost function', also
called `energy function'
\be
\label{energy}
H(t)=-\frac{1}{2} \sum_{i,j=1}^N J_{ij}(t)s_i(t)s_j(t) \, .
\ee
If $J_{ij}=J_{ji}$, eq. (\ref{energy}) is the central equation of the
Hopfield model \cite{hopfield}.
In case of an Ising system of atoms with spins, an equation of the
form (\ref{energy}) corresponds to the actual physical energy of the spin-system.

For a system of neurons, however, an energy function of the form
(\ref{energy}) is an ad-hoc postulate.
It is not derived from or suggested by some underlying biological,
biochemical or biophysical principle. 
In other words, the function (\ref{energy}), is, a priori, totally
unrelated to the actual energy of the neural system.
Consequently, a `derivation' leading to a Hebb rule based on a
function of the type (\ref{energy}), (see, e.g., \cite{hertz}), is just as ad hoc as the postulate underlying it.

In this section we will show that the Hebb rule
(\ref{the-one}) and its anti-Hebbian counterpart (\ref{other}) can be found by
postulating that the (\emph{biochemical}) \emph{energy} needed to change 
the synapses --- in order to store a new pattern $\XI$ --- is minimal.
We thus show that these particular Hebb rules --- and only these ones
--- are consistent with a physical principle.
The argument runs as follows.

The energy $\Delta E_{ij}$ to change the connection $w_{ij}(t_n)$ to
$w_{ij}(t_{n+1})$ will be a differentiable function of the magnitude of the change
$\Delta w_{ij}(t_n)$ occurring in (\ref{learning-step})
\be
\label{def-e}
\Delta E_{ij}=f_{ij}(\Delta w_{ij}) \, .
\ee
If a synapse between the neurons $i$ and $j$ is not changed in a learning step there is no
energy consumed.
Hence, the energy change $\Delta E_{ij}$ vanishes if $\Delta
w_{ij}=0$, i.e., 
\be
\label{e=0}
f_{ij}(0)=0 \, .
\ee
Moreover, we assume that a change of a synapse, whether it be a
strengthening or a weakening, can only be achieved by \emph{adding} energy to the
system. 
Thus, if $\Delta w_{ij} \neq 0$, we put,
\be
\label{e=pos}
f_{ij}(\Delta w_{ij}) > 0 \, .
\ee
The equations (\ref{e=0}) and (\ref{e=pos}) enable us to obtain a useful approximate expression for the energy
change $\Delta E_{ij}$.
We first note that any differential function $f(x)$ can
be written as a power series $f(x)=c^{(0)}+c^{(1)}x+c^{(2)}x^2+ \ldots$ \ .
Thus, we have for the function (\ref{def-e}), up to terms quadratic in
$\Delta w_{ij}$,
\be
\label{e-exp}
f_{ij}(\Delta w_{ij})=c_{ij}^{(0)}+c_{ij}^{(1)}\Delta
w_{ij}+c_{ij}^{(2)}\Delta w_{ij}^2 \, ,
\ee
where, in view of (\ref{e=0}) and (\ref{e=pos}) the coefficients have the properties 
\be
\label{c-fixed}
\begin{array}{lll} c_{ij}^{(0)}=0 \, , & c_{ij}^{(1)}=0 \, , & c_{ij}^{(2)} > 0
\end{array}
\ee
Furthermore, we take
\be
\label{c2}
c_{ij}^{(2)}=c_i \, ,
\ee
which is equivalent to the supposition that a change of connections
related to different
synapses $j=1,2,\ldots,N$ of the same neuron $i$ needs the same
amount of energy.
This assumption simplifies some of the formulae below; it is not
essential in the sense that all conclusions remain unaltered if the
simplification (\ref{c2}) is not used, see \cite{heerema}.
The total change $\Delta E$ in the $n$-th learning step $w_{ij}(t_n)
\longrightarrow w_{ij}(t_{n+1})$, where in principle all $w_{ij}$ with $j
\in V_i$ may change, is given by the sum of the individual changes, 
\be
\label{de-0}
\Delta E(\Delta w_{kl}) = \sum_{i=1}^N \sum_{j \in V_i} f_{ij}(\Delta
w_{ij}) \, ,
\ee
or, inserting (\ref{e-exp}) with (\ref{c-fixed}) and (\ref{c2}), by 
\be
\label{de}
\Delta E(w_{kl}(t_{n+1}))= \sum_{i=1}^N \sum_{j \in V_i} c_i \left(
w_{ij}(t_{n+1})-w_{ij}(t_n) \right)^2 \, . 
\ee
The positive constants $c_i$ are characteristic of neuron
$i$.

Equation (\ref{de}) will be our starting point for the derivation of
the energy saving learning rule (\ref{delta-lagrange}).
It is the general form any expression must have that describes the
energy needed to adapt the connection strengths $w_{ij}$ as a function
of their changes $\Delta w_{ij}$.
We now will minimize the change in energy $\Delta E$ as a function
of the new weights $w_{kl}(t_{n+1})$ under the
constraint (\ref{linear-fp}) using the Lagrange method.
This was the reason to write $\Delta E$ in (\ref{de}) as a function of
the $w_{kl}(t_{n+1})$ rather than as a function of
the $\Delta w_{kl}=w_{kl}(t_{n+1})-w_{kl}(t_n)$, as was done in (\ref{de-0}). 

\subsection{Storage of one pattern}
\label{p=1}

Let us consider at the $n$-th learning step, i.e., at time $t_n$, the storage of one pattern
$\XI$ in a network with connections given by $w_{ij}(t_n)$.
In case of a network with changing and non-changing weights as introduced in
section~\ref{network}, the expression for the change of energy is,
up to second order in the changes of the synaptic weights, given by
(\ref{de}).
Note that a minimization of the one condition
(\ref{de}) under the constraint induced by the fixed point equation
(\ref{linear-fp}) implies a minimization of the $N^2-M$ changes $\Delta
w^2_{ij}(t_n)$, since a sum of positive terms is minimal if and only if
each term is minimal; recall that $M$ is the number of synapses with
constant weights $w_{ij}$.

For the storage of one single pattern $\XI$, one may rewrite the fixed point equations
(\ref{linear-fp}) in the form
\be
\label{g=0}
g_i(w_{ij}(t_{n+1}))=0 \, , \quad (i=1,\ldots,N) \, ,
\ee
where
\be
\label{g}
g_i(w_{ij}(t_{n+1}))=\kappa(2\xi_i-1)- \sum_{j \in V_i^c}
w_{ij}(t_n)\xi_j - \sum_{j \in V_i} w_{ij}(t_{n+1})\xi_j+ \theta_i
\, . 
\ee
The method of Lagrange multipliers tells that one finds the extrema of
(\ref{de}) subject to the auxiliary conditions (\ref{g=0}) from the
$N^2-M$ equations 
\be
\label{f+g}
\frac{\partial \Delta E}{\partial w_{ij}(t_{n+1})}+\sum_{k=1}^N  \lambda_k
\frac{\partial g_k}{\partial w_{ij}(t_{n+1})}=0 \, , \quad
(i=1,\ldots,N; j \in V_i) \, ,
\ee
Upon substitution of (\ref{de}) and (\ref{g}) into this expression, we
find the $N^2-M$ relations
\be
\label{solution-f+g}
w_{ij}(t_{n+1})=w_{ij}(t_n)+\frac{1}{2c_i} \lambda_i\xi_j \, , \quad
(i=1,\ldots,N; j \in V_i) \, .
\ee 
In the method of Lagrange multipliers the number of constraints
equals the number of Lagrange multipliers $\lambda_i$.
Hence, there are $N$ Lagrange multipliers.
Since the $N$ multipliers $\lambda_i$ are unequal
to zero, it follows from the $N^2-M$ equations (\ref{solution-f+g})
that $N^2-M \ge N$, or $M \le N^2-N$.
We now have obtained the $N+N^2-M$ equations (\ref{g=0}) and
(\ref{solution-f+g}) for the $N+N^2-M$ unknowns $\lambda_i$ and $w_{ij}(t_{n+1})$.

The structure of these equations happens to be such that an explicit
expression for the $\lambda_i$'s can be found, and thereupon, an
explicit expression for the $w_{ij}(t_{n+1})$'s can be obtained.
The procedure is as follows.

Eliminating the $w_{ij}(t_{n+1})$'s from (\ref{g=0}) with the help of
(\ref{solution-f+g}), leads to
\be
\label{lambda-1}
\lambda_i= \frac{2c_i}{\sum_{k \in V_i} \xi_k} \left[
\kappa-\gamma_i(t_n) \right] (2\xi_i-1) \, ,
\ee
where we used the property $(\xi_j)^2=\xi_j$.
Substituting this expression for $\lambda_i$ into
(\ref{solution-f+g}) yields
\be
\label{w-lagrange-1}
w_{ij}(t_{n+1}) = w_{ij}(t_n) +\frac{1}{\sum_{k \in V_i} \xi_k} \left[ \kappa -
\gamma_i(t_n) \right] (2\xi_i-1) \xi_j \, ,
\quad (j \in V_i) \, ,
\ee
or, equivalently [see eq. (\ref{learning-step})],
\be
\label{delta-lagrange}
\Delta w_{ij}(t_n)=\frac{1}{ \sum_{k \in V_i} \xi_k} \left[ \kappa -
\gamma_i(t_n) \right] (2\xi_i-1) \xi_j \, , \quad
(j \in V_i) \, ,
\ee
where $\kappa$ is the positive parameter (\ref{stable}) related to the basins of
attraction, and where the $\gamma_i$ ($i=1,\ldots,N$) are the
stability coefficients given by (\ref{gamma}).
We will refer to (\ref{delta-lagrange}) by the name of \emph{non-local
  energy saving learning rule}, since the denominator of (\ref{delta-lagrange})
 depends on the input from all neurons $k$ that are connected via
 changing connections to neuron $i$.
The factor between square brackets
\be
\label{k-g}
 \kappa -
\gamma_i(t_n)= \kappa - ( h_i(t_n) -\theta_i) (2
\xi_i-1) 
\ee
depends solely upon the temporal and environmental state of the post-synaptic neuron $i$, that
is, on its post-synaptic potential $h_i$ at time $t_n$ of the $n$-th
learning step, its thresholds
$\theta_i$, its activity $\xi_i$ and a parameter $\kappa$.
The factor (\ref{k-g}) can be positive or negative.
Therefore, the learning rule (\ref{delta-lagrange})--(\ref{k-g})
derived here from the assumption of minimal energy change per
learning step, happens to coincide with the particular Hebbian
learning rule (\ref{the-one}) and its anti-Hebbian counterpart
(\ref{other}) found in section~\ref{hebb} on purely intuitive grounds,
grounds which were related to biological plausibility.

We thus have shown that if biological neurons would adapt their connections
according to the non-local energy saving learning rule (\ref{delta-lagrange}), this adaptation
would be such that the network would fulfil the fixed point equation
(\ref{linear-fp}) for a pattern $\XI$.
Moreover, the learning rule (\ref{delta-lagrange}) guarantees that the energy needed to rebuild a neural network with
connections $w_{ij}(t_n)$ to a network with connections
$w_{ij}(t_{n+1})$ is minimal.

We conclude this section with some remarks.
The energy saving learning rule is only applicable in
those situations in which the denominator is unequal to zero.
This can be translated into a restriction on the $\xi_k$, $k \in V_i$.
It follows that with an decreasing number of adaptable connections
there is an increasing number of patterns that cannot be stored with the help of the
non-local energy saving learning rule. 
This effect will be absent when the local energy saving learning rule
is used (see section~\ref{local}). 

When we repeat the derivation of (\ref{delta-lagrange}) in the
spin-representation with time-dependent thresholds as given by
(\ref{def-t}), we find again (\ref{delta-lagrange}) with $\xi$
replaced by $(s+1)/2$, i.e.,
\be
\label{spin-binary}
\Delta J_{ij} \propto s_i (s_j+1) \, , 
\ee
as could be expected.
If, however, the derivation of (\ref{delta-lagrange}) is repeated in
the spin representation with $T_i$ taken to be a constant, as is
usually done in the spin-representation, one finds a result which
differs from (\ref{spin-binary}), namely 
\be
\label{spin-spin}
\Delta J_{ij} \propto s_i s_j\, .
\ee
This is the biologically less relevant result commonly encountered in
the physical literature, as noticed already in section~\ref{hebb}: see
eq. (\ref{sym}).

\subsection{Storage of $p$ patterns}
\label{p>1}

In the previous section we saw that storage of one pattern $\XI$
can be achieved via a synaptic change $\Delta w_{ij}$ given by
(\ref{delta-lagrange}). 
Hence, storage of $p$ patterns $\XI^{\mu}$ ($\mu=1,\ldots,p$) might be
accomplished by repeated application of the learning rule
(\ref{delta-lagrange}).
Let us therefore consider the following learning process.
In a first interval of time, $[t_0,t_1)$, a first pattern $\XI^1$ is stored
via the change $\Delta w_{ij}(t_0)$, leading to the connections
$w_{ij}(t_1)=w_{ij}(t_0)+\Delta w_{ij}(t_0)$, $j \in V_i$.
Next, in the interval $[t_1,t_2)$, pattern $\XI^2$ is stored, etcetera.
Finally, pattern $\XI^p$ is stored.
We call this sequence of storage of $p$ patterns a learning cycle.

The energy saving learning rule is a storage prescription for a new
pattern, which does not take into account, however, any constraint
that would guarantee that a previously stored patterns remain stored.
Thus it may occur that storage of a new pattern will perturb,
partially or totally, the storage of an older pattern.

In section~\ref{ideal}, on maximal learning efficiency, we will
determine a learning rule which does guarantee that new patterns
are stored without wiping out previously stored patterns.
However, this learning rule will turn out to be biologically
unacceptable.
We therefore proceed with the learning rule derived above.
We shall derive, along the lines of reasoning of Diederich and Opper
\cite{opper}, but for diluted networks, an expression for the
weights $w_{ij}$ of the synaptic connections after infinitely many
learning cycles.
It will turn out that, in the end, previously stored patterns are not
forgotten.
 
As follows from eq. (\ref{learning-step}), the connections after $R$
learning cycles are given by 
\be
\label{r-cycle}
w_{ij}(t_{Rp})= w_{ij}(t_0)+\sum_{m=1}^R
\sum_{\mu=1}^p \Delta w_{ij}(t_{(m-1)p+\mu-1}) \, , \quad (j \in V_i) \, ,
\ee
with $t_{Rp}$ the time after $R$ learning cycles of $p$ patterns.

Substituting (\ref{delta-lagrange}) into (\ref{r-cycle}) we find
\be
\label{w}
w_{ij}(t_{Rp})=w_{ij}(t_0)+N^{-1} \sum_{\mu=1}^p F_i^{\mu}(t_{(R-1)p+\mu-1})\xi_j^{\mu} \, , \quad (j \in
V_i) \, ,   
\ee
where 
\bea
\label{e}
F_i^{\mu}(t_{(R-1)p+\mu-1}) &=& \sum_{m=1}^R [ \kappa(2\xi_i^{\mu}-1)
-  (\sum_{k \in V_i^c}  w_{ik}(t_0)\xi_k^{\mu} \nn\\
& & + \sum_{k \in V_i} w_{ik}(t_{(m-1)p+\mu-1})\xi_k^{\mu}-\theta_i )
] / (N^{-1} \sum_{k \in V_i} \xi_k^{\mu}) 
\eea
is the effect on $w_{ij}$ of pattern $\XI^{\mu}$ after $R$ learning
cycles have been completed.
From (\ref{e}) it follows that
\bea
\label{de-1}
\lefteqn{ (N^{-1} \sum_{k \in V_i} \xi_k^{\mu}) [ F_i^{\mu}(t_{(R-1)p+\mu-1})-F_i^{\mu}(t_{(R-2)p+\mu-1})] =
\kappa(2\xi_i^{\mu}-1) } \nn\\
&& - (\sum_{k \in V_i^c} w_{ik}(t_0)\xi_k^{\mu} + \sum_{k \in V_i}
w_{ik}(t_{(R-1)p+\mu-1})\xi_k^{\mu}  -\theta_i ) \,  .
\eea

In the $R$-th learning cycle, at time $t_{(R-1)p+\nu-1}$, only the
patterns $\XI^{1},\ldots,\XI^{\nu -1}$ have changed the weights of the
network.
Hence, the $F_i^{\nu}$ with $\nu<\mu$ have new values at time
$t_{(R-1)p+\mu-1}$, whereas the $F_i^{\nu}$ with $\nu \ge \mu$ are
still identical to their values in the previous learning cycle, i.e.,
are equal to the values at time $t_{(R-2)p+\mu-1}$.
Thus, with the help of (\ref{w}), the weights in the right-hand side
of (\ref{de-1}) can be expressed as follows in terms of the $F_i^{\mu}$:
\be
\label{de-2}
w_{ik}(t_{(R-1)p+\mu-1})= w_{ik}(t_0)+ N^{-1}\sum_{\nu<\mu}
F_i^{\nu}(t_{(R-1)p+\mu-1})\xi_k^{\nu} + N^{-1} \sum_{\nu \ge \mu}
F_i^{\nu}(t_{(R-2)p+\mu-1})\xi_k^{\nu} \, . 
\ee
Eliminating $w_{ik}(t_{(R-1)p+\mu-1})$ from (\ref{de-1}) with the help
of (\ref{de-2}) yields 
\bea
\label{gauss}
\lefteqn{ N^{-1} \sum_{k \in V_i} \sum_{\nu \le \mu}
  F_i^{\nu}(t_{(R-1)p+\mu-1}) \xi_k^{\nu}\xi_k^{\mu} =  - N^{-1}
  \sum_{k \in V_i} \sum_{\nu > \mu} F_i^{\nu}(t_{(R-2)p+\mu-1})
  \xi_k^{\nu}\xi_k^{\mu} }\nn\\
&& + [\kappa -\gamma_i^{\mu}(t_0)](2\xi_i^{\mu}-1) \, . 
\eea
This system of linear equations can be solved for $F_i^{\mu}$ using the Gauss-Seidel
iterative method.
We first rewrite (\ref{gauss}) in matrix
notation.
Next, we introduce a $p \times p$ matrix $C_i$,
the matrix-elements of which are given by 
\be
\label{c}
C_i^{\mu \nu}:=N^{-1} \sum_{k \in V_i} \xi_k^{\mu}\xi_k^{\nu} \, .
\ee
We might call this matrix the `reduced correlation matrix', since it
correlates $\xi_k^{\mu}$ and $\xi_k^{\nu}$ while taking into account,
via $V_i$, the connectivity of the network.
The reduced correlation matrix is closely related to the usual correlation
matrix if $V_i$ contains all neuron indices. 
We proceed by decomposing this matrix $C_i$ into matrices $L_i$ and
$U_i$ in such a way that $C_i=L_i+U_i$.
The matrix $L_i$ is a matrix with only non-zero matrix-elements
on and below the diagonal and $U_i$ is a matrix with only non-zero
matrix-elements above the diagonal.
We also introduce the vectors ${\bf F}_i(R):=(F_i^1(t_{(R-1)p+1-1)}), \ldots,
F_i^p(t_{(R-1)p+p-1)})) $ and ${\bf G}_i:=([\kappa
-\gamma_i^1(t_0)](2\xi_i^1-1), \ldots ,[\kappa -\gamma_i^p(t_0)](2\xi_i^p-1))$.
Finally, we shall denote a $p \times p$ unit matrix as $I$.
We thus can rewrite (\ref{gauss}) in the form
\be
L_i \cdot {\bf F}_i(R)= -U_i \cdot {\bf F}_i(R-1)+ {\bf G}_i.
\ee
By iteratively solving this equation for ${\bf F}_i(R)$, we find 
\bea
\label{e-r}
\lefteqn{ {\bf F}_i(R) = \left[ - L_i^{-1} \cdot U_i \right]^{R-1}
  \cdot {\bf F}_i(1) } \nn\\
& & + L_i^{-1} \cdot \left[I - L_i^{-1} \cdot U_i +
  \cdots + (-L_i^{-1} \cdot U_i)^{R-2} \right] \cdot {\bf G}_i \, .
\eea
The symmetric matrix $C_i$, as defined in (\ref{c}), is
  positive definite and symmetric.
It then can be shown that the matrix $ - L_i^{-1} \cdot U_i$ has
  eigenvalues smaller than one \cite{seidel}.
As a consequence, we have 
\be
\label{sym-posdef}
\lim_{R \rightarrow \infty} \left[ - L_i^{-1} \cdot U_i \right]^{R-1}=0,
\ee 
and it follows that, in the limit $R \rightarrow \infty$, (\ref{e-r})
converges to
\bea
\label{e-final}
{\bf F}_i(\infty) &=& L_i^{-1} \cdot \left[ I-(-L_i^{-1} \cdot U_i)
\right]^{-1} \cdot {\bf G}_i \nn\\
&=& C_i^{-1} \cdot {\bf G}_i,
\eea
where ${\bf F}_i(\infty)=\lim_{R \rightarrow \infty} {\bf
  F}_i(R)$.
Substitution of (\ref{e-final}) in (\ref{w}) and restoring the old
  notation, yields, for $R \rightarrow \infty$
\be
\label{w-opper}
w_{ij}(t_{\infty}) = \left\{ \begin{array}{ll} {\displaystyle
  w_{ij}(t_0)+ N^{-1} \sum_{\mu,\nu=1}^p \left[\kappa
  -\gamma_i^{\mu}(t_0) \right] (2\xi_i^{\mu}-1) (C_i^{-1})^{\mu
  \nu} \xi_j^{\nu} } \, , & (j \in V_i) \\
w_{ij}(t_0) \, , & (j \in V_i^c) \end{array} \right.
\ee
where $(C_i^{-1})^{\mu \nu}$ is the inverse of the matrix (\ref{c}).

By substituting (\ref{w-opper}) into (\ref{linear-fp}) it can directly
be verified that the weights (\ref{w-opper}) fulfil (\ref{linear-fp})
for all $\mu$ ($\mu=1,\ldots,p$).
For $p=1$ this was to be expected, since the learning rule
(\ref{delta-lagrange}) was constructed that way.
For $p>1$ one could, for the same reason, expect that (\ref{linear-fp})
would be verified by (\ref{w-opper}) for the final pattern of the
learning cycle, $\XI^p$.
It is less transparent, however, that (\ref{w-opper}) satisfies
(\ref{linear-fp}) for all patterns $\XI^{\mu}$.

The result (\ref{w-opper}) is exact for networks
with a number of vanishing connections running from $M_0=0$ to
$M_0=N^2-Np$, i.e., valid for dilution $0$ to $d=1-\alpha$, where
$\alpha=p/N$.
The analogous calculation performed by Diederich and Opper for networks
with empty $V_i^c$, so that $V_i$ contains all indices, yields a
result that coincides with
the result obtained via the usual pseudo-inverse solution
\cite{personnaz,kohonen} of eq. (\ref{linear-fp}).
Hence, the following question may now arise. 
Can we solve the eq. (\ref{linear-fp}) for a neural network where
$V_i^c$ is not empty and, consequently, the method of the pseudo-inverse in its
standard form is not applicable?
The answer to this question is affirmative. 
In \ref{pseudo-inverse} we modify the method of the pseudo-inverse so as to be
applicable to systems with changing and non-changing interactions.
Solving eq. (\ref{linear-fp}) for networks with changing and
non-changing connections via what we have called the modified method
of the pseudo-inverse, one indeed obtains (\ref{w-opper}), as we also prove in the appendix.

Thus we have shown that the solution that corresponds to the stepwise
  energetically most economic way to realize storage of patterns in a
  partially connected network, turns out to be identical to the one obtained 
  via a ---modified--- version of the well-known mathematical method of the
  pseudo-inverse applied to the fixed point equation (\ref{linear-fp}).
In other words, the non-local energy saving learning rule (\ref{delta-lagrange}) leads
  to the solution of the fixed point equation (\ref{linear-fp}),
  obtained via the modified method of the pseudo-inverse, which is
  based, in turn, on the reduced correlation matrix.

We conclude this section with a few remarks.
 In general, the inverse of the matrix $C_i^{\mu \nu}$ cannot easily
 be found analytically.
However, in the non-biological case that none of the weights is kept constant, all index sets $V_i^c$ are
  empty.
As a consequence one may use, for large $N$ and low storage capacity
 $\alpha:=p/N$, the approximations
\bea
\label{N-oneindig-1}
N^{-1}\sum_{j=1}^{N}\xi_j^{\mu} &=& a \\
\label{N-oneindig-2}
N^{-1}\sum_{j=1}^{N}\xi_j^{\mu}\xi_j^{\nu} &=& a^2 \, , \quad (\mu \neq \nu)
\eea
Substitution of (\ref{N-oneindig-1}) and (\ref{N-oneindig-2}) into
(\ref{c}), where now $V_i$ is the set of all indices, yields 
\be
\label{c-approx}
C_i^{\mu \nu}=a(1-a)\delta_{\mu \nu}+a^2 \, .
\ee
For the inverse of $C_i^{\mu \nu}$ we thus obtain from
(\ref{c-approx}) the simple analytical expression
\be
\label{c-inverse}
(C_i^{-1})^{\mu \nu}=\frac{1}{a(1-a)} \left[ \delta_{\mu
    \nu}-\frac{a}{ap-a+1} \right] \, .
\ee
Using (\ref{c-inverse}) in (\ref{w-opper}), leads to
\bea
\label{w-lagrange}
\lefteqn{w_{ij}(t_\infty)=w_{ij}(t_0)-
\frac{1}{Na(1-a)}\frac{a}{ap-a+1} \sum_{\mu,\nu=1}^p \left[
\kappa -\gamma_i^{\mu}(t_0) \right] (2\xi_i^{\mu}-1)
\xi_j^{\nu} } \nn\\
& & +\frac{1}{Na(1-a)} \sum_{\mu=1}^p \left[ \kappa - \gamma_i^{\mu}(t_0)
\right] (2\xi_i^{\mu}-1) \xi_j^{\mu} \, , \quad (i,j=1,\ldots,N)  \, .
\eea
Equation (\ref{w-lagrange}) is an explicit expression for the weights
$w_{ij}$ of a (non-biological) network in which all the weights,
including the self-interactions $w_{ii}$, are present.

Kanter and Sompolinsky used the result (\ref{w-opper}) in case $i
\neq j$ for a fully connected network without self-interactions \cite{kanter}.
Their ad-hoc assumption that the self-interactions $w_{ii}$ can be put
equal to zero, turns out to be justified in view of our exact result
(\ref{w-opper}) with $w_{ii}(t_0)=0$.

\section{A learning rule with maximal learning efficiency}
\label{ideal}

In the preceding section learning of a collection of patterns was
achieved by repeated application of the non-local energy saving learning rule.
This learning rule was not constructed in such a way that conservation of
storage of old patterns was automatically guaranteed when a new pattern was stored.
We now address the question whether and how storage of a new pattern
$\XI^{p+1}$ can be achieved without disturbing the storage of the old
patterns $\XI^1,\ldots,\XI^p$.
We shall refer to this type of learning as maximally efficient learning.

Linkevich \cite{linkevich} treated this problem on the basis of a mathematical
model, in which suppositions are made which cannot be true in a
biological neural network.
Firstly, he treated the thresholds $T_i(t)$, eq. (\ref{def-t}), as a
vanishing constant.
Moreover, his network has symmetric
connections $w_{ij}(t)=w_{ji}(t)$, whereas a biological network has
non-symmetric connections $w_{ij}(t) \neq w_{ji}(t)$.
Finally, his network is fully connected, i.e., all
$w_{ij}(t)\neq 0$.

We may improve and generalize the reasoning of Linkevich to obtain a maximally efficient
learning rule for a partially connected network with non-symmetric
connections.
The calculations only hold for networks in which the
thresholds are equal to the stability
coefficients $\kappa$, i.e., $\theta_i=\kappa$, for all $i$, and in case the initial
connections are equal to zero, $w_{ij}(t_0)=0$ for all $i$ and $j$.
As a final result we arrive, in this particular case, at the following
rule for learning with maximal learning efficiency (see \ref{efficient})
\be
\label{dw-ideal}
\Delta w_{ij}(t) = \frac{{\displaystyle [ \kappa -\gamma_i^{p+1}(t) ]
    [ \kappa- \gamma_j^{p+1}(t) ] (2\xi_i^{p+1}-1) (2\xi_j^{p+1}-1)}}
{{\displaystyle \sum_{l \in V_i}  [ \kappa - \gamma_l^{p+1}(t) ]
    \xi_l^{p+1} }} \, , \quad  (j \in V_i) \, .
\ee

From (\ref{dw-ideal}) we immediately see that, in general, $\Delta
w_{ij}$ is not symmetric in $i$ and $j$.
However, for a network in which all connections may change we find that $\Delta
w_{ij}$ is symmetric in $i$ and $j$, in accordance with the result of
Linkevich.
Note that the $i$- dependent factors in the numerators of
(\ref{dw-ideal}) and (\ref{delta-lagrange}) are identical, which reflects
the fact that the new pattern $\XI^{p+1}$ has to obey the fixed point
equation, both in the cases of `stepwise minimal change in energy'
(\ref{delta-lagrange}) and of `stepwise maximal efficient learning' (\ref{dw-ideal}).

The learning rule with maximal learning efficiency (\ref{dw-ideal}) is of the
form (\ref{sym}), a form
which we have rejected, in section~\ref{hebb}, on biological grounds.
We therefore shall not pursue any further the analysis of the learning rule with
maximal learning efficiency in the remainder of this article.

\section{Locality of learning rules}
\label{local}

Up to now we did not mention an important limitation of a biological
learning rule.
The mathematical learning rule to change a weight of a network can, in
principle, be local or non-local.
The second possibility must be excluded in case the weight is associated with a synapse: there is no biological
construction available in the brain to tell a specific synapse how and when to
change as a function of properties of neurons with which it has no
direct contact.
The modifications must result from the \emph{local} situation, i.e.,
limited to the situation spatially `close enough' to the synapse in
question, and within a `brief span' of time.
Thus, a change $\Delta w_{ij}$ may depend only on variables local, in
space and time, to the neurons $i$ and $j$.
The local variables available at the synapse between neurons $i$ and $j$
are the activities $\xi_i$ and $\xi_j$, the post-synaptic potentials $h_i$ and
$h_j$, and the thresholds $\theta_i$ and $\theta_j$.
Hence, the factors $\epsilon_{ij}$ occurring in
Hebb rules should depend on these variables only
\be
\epsilon_{ij}=\epsilon_{ij}(\xi_i,h_i,\theta_i,\xi_j,h_j,\theta_j) \, .
\ee

The energy saving learning rule (\ref{delta-lagrange}) for $\Delta w_{ij}$
guarantees, after repeated application, storage of patterns in a way
which is energetically efficient.
The factor between square brackets in the non-local learning rule (\ref{delta-lagrange}) fulfils the criterion
of locality.
However, the learning rule as a whole is not a local learning rule because of the factor,
\be
\label{factor}
1/ \sum_{k \in V_i} \xi_k
\ee
which depends, because of the sum over $k$'s restricted to $V_i$,
eq. (\ref{v}),  on the network connectivity, and hence, not on properties
related to neurons $i$ and $j$ only.
If we approximate (\ref{factor}) by some constant, $\eta_i$ say, we do
obtain a learning rule that is local,
\be
\label{dw-local}
\Delta w_{ij}(t_n)=\eta_i \left[ \kappa - \left( h_i(t_n)
-\theta_i \right) (2\xi_i-1) \right] (2\xi_i-1) \xi_j \, .  
\ee
We shall refer to (\ref{dw-local}) as the \emph{local energy saving
  learning rule}.  
The better $\eta_i$ approximates a value dictated by (\ref{factor}),
the better this local learning rule will be with respect to its
energetic efficiency.

At this point it is important to note that the proof of convergence of
section~\ref{p>1} can be generalized, replacing everywhere the factor
(\ref{factor}) by the constant positive factor $\eta_i$.
As a final result (\ref{w-opper}) is found again, provided certain
restrictions on $\eta_i$ are satisfied.
It then can be proved \cite{heerema} that the local, biologically realizable energy
saving learning rule yields the same final values $w_{ij}(t_{\infty})$ as the non-local
energy saving learning rule.

As noticed in section~\ref{introduction}, the constant $\eta_i$ is a neuron property, the
determination of which is outside the scope of the present article: we
then would have to determine the coefficients $c_{ij}$ in the
expression for $f_{ij}$ (\ref{e-exp}) explicitly.

A reasonable approximation for $\eta_i$ can easily be obtained for a
fully connected network where all connections may change in time.
For such a network we have the approximation (\ref{N-oneindig-1}) for
the denominator of (\ref{factor}), 
which implies
\be
\label{eta}
\eta_i \approx (Na)^{-1} \, , \quad {\rm for \ all} \ i=1,\ldots,N \, .
\ee 
We will use this approximation in the following section where we
consider a biological network.

\section{Local versus non-local learning}
\label{numerical}

In this section, we will study numerically, for a biological network
with dilution $d$, the local energy saving learning rule
(\ref{dw-local}) as a competitor of the non-local learning rule
(\ref{delta-lagrange}).
For $\eta_i$ we take, quite arbitrarily, the constant (\ref{eta}).
We could as well have taken $1/N$ or $1/(N(1-d))$: the essentials of
the behaviour of the numerical results are not very sensitive for the
precise values of the $\eta_i$.

In order to judge the functioning of a recurrent network with respect
to its ability to store an arbitrary collection of $p$ patterns
$\XI^{\mu}$ ($\mu=1,\ldots,p$), we take $L$ sets of such collections,
and label them by $\XI^{\mu,m}$ ($m=1,\ldots,L$), i.e., $\XI^{\mu,m}$ is pattern $\mu$ of set $m$.
The performance of the network with respect to the patterns from the $m$-th set may be characterized by the $Np$
stability coefficients $\gamma_i^{\mu,m}$ ($i=1,\ldots,N$;
$\mu=1,\ldots,p$) defined in equation (\ref{gamma}).
The stability coefficients $\gamma_i^{\mu,m}$ should be positive
[see eq. (\ref{fp-equiv})]. 
Moreover, we have normalized in such a way that the $\gamma$'s should
be close to one.
Hence, the more $\gamma_i^{\mu,m}$ we find with values around one,
the better the network will perform.

We first define for the particular set $m$ of $p$ patterns the quantity:
\be
\gamma^{m}={\displaystyle {\rm min}_{i=1,\ldots,N} } \left\{
\gamma_i^{1,m},\ldots,\gamma_i^{p,m} \right\} \, .
\ee
Hence, $\gamma^{m}$ is the minimal value of all stability
coefficients for a particular set $m$ of $p$ patterns.
A network does not function if $\gamma^{m}$ is negative, and
functions better and better when $\gamma^{m}$ becomes closer to one
(with the normalization $\kappa=1$).
To find a number that characterizes the network performance for an
arbitrary set of $p$ patterns, we average the minimal values
$\gamma^{m}$ over $L$ arbitrarily chosen sets,
\be
\gamma=\frac{1}{L} \sum_{m=1}^L \gamma^{m} \, .
\ee
Hence, $\gamma$ is the average with respect to the $L$ sets of $p$
patterns $\xi^{\mu}$.
We therefore will refer to $\gamma$ as the average performance of the
network.
Similarly, we define the average energy change $\Delta E$
\be
\label{de-av}
\Delta E=\frac{1}{L} \sum_{m=1}^L \Delta E^{m} \, ,
\ee
where $\Delta E^{m}$ is the change of energy in one learning step of the $m$-th set of patterns.
Furthermore, we define the average energy change per synapse $\Delta
e$, as
\be
\label{de-syn}
\Delta e=\Delta E/ (N^2-M) \, ,
\ee 
where $M$ is the number of non-changing synapses.
We also will study the performance of
neural networks with varying dilution by considering the distribution
of the stability coefficients $\gamma_i^{\mu,m}$.
By studying numerically the quantities $\gamma$ and $\Delta e$ and the
distribution of the stability coefficients $\gamma_i^{\mu,m}$, we
can judge the power of the (exact) non-local energy saving learning rule (\ref{delta-lagrange})
compared to the (biologically feasible) local energy saving learning rule (\ref{dw-local})--(\ref{eta}).

\subsection{Storage of one pattern}

\begin{figure}[htb]
\centerline{\hbox{\epsffile{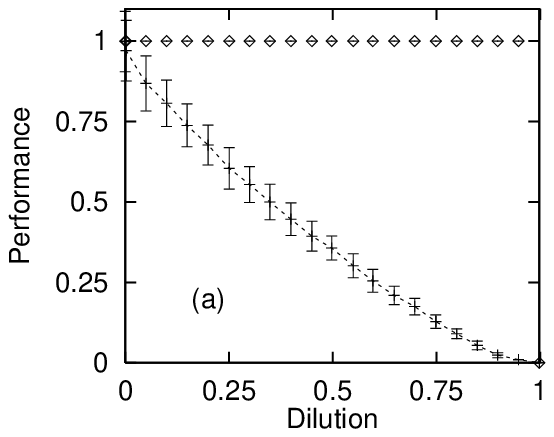} \epsffile{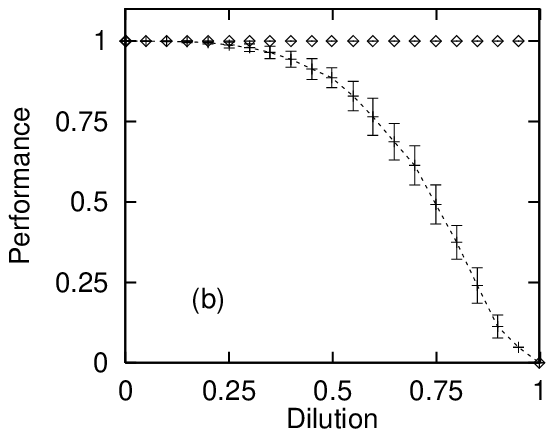}}}
\centerline{\hbox{\epsffile{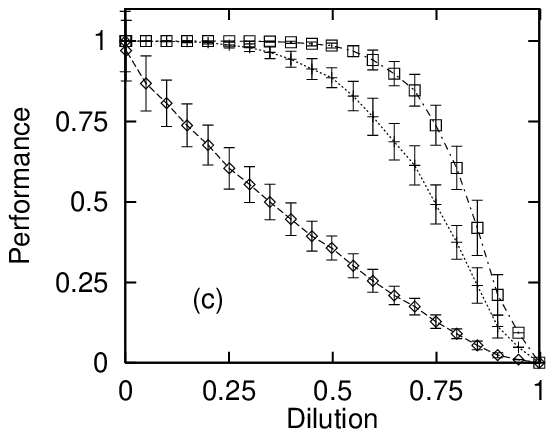}}}
\protect\caption[1]{
The average performance, $\gamma$, of a network of $512$ neurons as a
function of its dilution $d$.
Dilution $d=0$ means that the network is fully interconnected ($w_{ij}
\neq 0$ for all $i$ and $j$), dilution $d=1$ means that there are no
connections anymore ($w_{ij}=0$ for all $i$ and $j$).
The one pattern $\XI$ is chosen arbitrarily, but
such that the mean activity $a=0.2$.
The computations have been averaged over $100$ different $\XI$.
The error bars give the standard deviation of the averaged stability 
coefficients $\gamma_i$ ($i=1,\ldots,N$).
The calculations are performed starting from a tabula rasa for the
weights ($w_{ij}(t_0)=0$) and vanishing thresholds ($\theta_i=0$). 
\protect\\
Figures (a),(b).
In the first two figures, a comparison between
  the non-local energy saving learning rule (\ref{delta-lagrange}) (upper
  curves) and the local energy saving learning rule (\ref{dw-local})
  (lower curves) after it has been applied one, (a), and five, (b), times. 
\protect\\
Figure (c).
In the last figure, a comparison of the local energy saving learning rule
(\ref{dw-local}) after it has been applied one (lower curve), five and ten (upper curve) times.
\protect\label{fig1}}
\end{figure}

\paragraph{Performance}
The non-local energy saving learning rule (\ref{delta-lagrange}) and its local
approximation (\ref{dw-local})--(\ref{eta}) are used to store one
pattern $\XI$.
In order to compare the quality of the two learning rules we have
plotted in figure~\ref{fig1} the average performance $\gamma$ versus
the dilution $d$ of the network for both learning rules.
We see that the non-local learning rule stores a new pattern such
that $\gamma=1$, as could be expected since it has been designed that
way.
Moreover, we see that both the non-local and the local learning rules
lead to positive values of $\gamma$, and, hence, lead to storage of
the pattern $\XI$.
The non-local learning rule, however, leads at once to $\gamma=1$,
whereas the local learning rule converges to $\gamma=1$ only after repeated
application.
Hence, basins of attractions of the local learning rule are smaller
initially [see figure~\ref{fig1}].

\paragraph{Use of energy}
Furthermore, we consider the average energy change per synapse $\Delta e$ (\ref{de-syn}) for the
non-local and local learning rules as a function of the number of
synapses in a network of a fixed number of neurons.
In case of a single application of an energy saving learning rule, it
turns out that for the non-local learning rule $\Delta e$ increases as
the number of synapses decreases, while $\Delta e$ is constant in case
of the local learning rule.
This favourable situation of remaining constant apparently is an unexpected positive effect
of the approximation made when going from a non-local energy saving
learning rule to a local energy saving learning rule.

\begin{figure}[htb]
\centerline{ \hbox{\epsffile{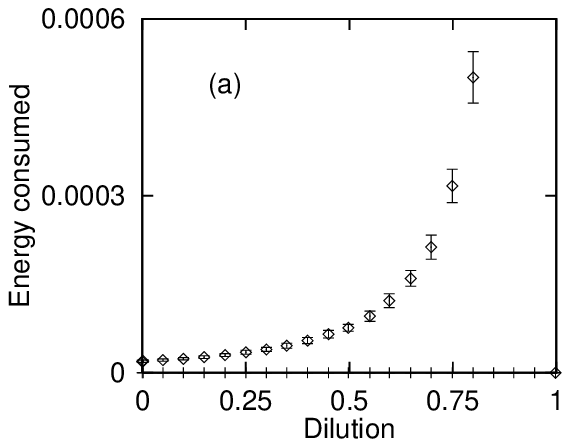} \epsffile{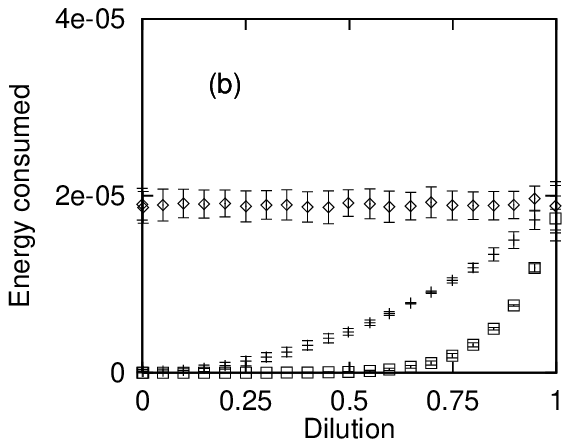}} }
\protect\caption[2]{The average energy consumed per synapse $\Delta e$ in one
  learning step, of a network of $512$ neurons as a function of its
  dilution $d$.
The one pattern $\XI$ is chosen arbitrarily, but
such that the mean activity $a=0.2$.
The computations have been averaged over $100$ different $\XI$.
The error bars give the standard deviation of the averaged stability 
coefficients $\gamma_i$ ($i=1,\ldots,N$).
The calculations are performed starting from a tabula rasa for the
weights ($w_{ij}(t_0)=0$) and vanishing thresholds ($\theta_i=0$). 
\protect\\
Figure (a).
The average energy change per synapse $\Delta e$ for the non-local energy
saving learning rule after one (upper curve) and two learning steps
(lower curve, coinciding with the horizontal axis).
\protect\\
Figure (b).
The average energy change per synapse $\Delta e$ for the local energy saving
learning rule caused by the first (upper curve), second or fifth
(lower curves) time that the local energy saving rule
(\ref{dw-local})--(\ref{eta}) is used.
\protect\label{fig2} }
\end{figure}

In case of repeated application there almost is no energy effect for
the non-local learning rule, and a slight effect for the local learning
rule: the energy need per synapse grows with growing dilution [see
figure~\ref{fig2}].

\subsection{Storage of $p$ patterns}

Having studied numerically the storage of one pattern, we now turn to
the storage of $p$ patterns. 
As pointed out in section~\ref{p>1} this may be achieved through
repeated application of the energy saving learning rule.

Storage of one pattern ($p=1$) could be achieved in such a way that, by
construction, all $\gamma_i^{\mu, m}$ ($\mu=1$) were equal to one in case of the
non-local learning rule: $\gamma_i^{1, m}=1$ for all $i$ and $m$.
As a consequence, the local energy saving learning rule, which is an approximation
to the non-local one, has the property that all $\gamma_i^{1, m}$ are
`not too far away' from the value $\kappa=1$, i.e., they are positive.
We recall that positivity of the stability coefficients $\gamma_i^{1,m}$ is a sufficient criterion for a network to store what should be
  stored [see figure~\ref{fig1}].

When the energy saving learning rule is used to store more than one pattern, the
positivity of all but the last stored pattern is not guaranteed.
As noted before, we must allow for the fact that storage of a new pattern may spoil the storage of older
patterns.
Therefore, the requirement that the minimum of all $\gamma_i^{\mu,m}$
($\mu=1,\ldots,p$) should be positive is too strong.
Forgetting thus turns out to be an inevitable consequence of storing new
patterns, at least in the beginning.
By repeating the learning procedure for whole sequences of patterns we
can achieve that more and more $\gamma_i^{\mu,m}$ become positive,
suggesting that more and more patterns may be definitely stored.

\begin{figure}[htb]
\centerline{\hbox{\epsffile{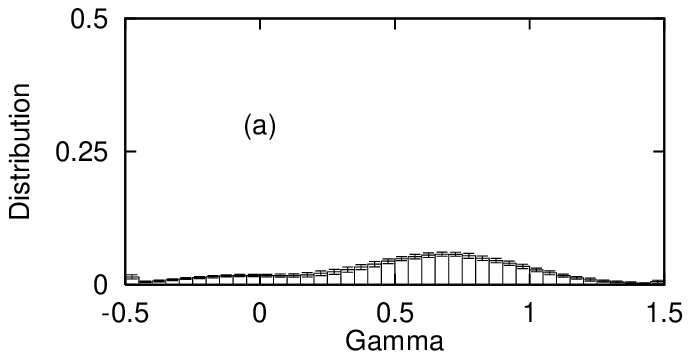} \epsffile{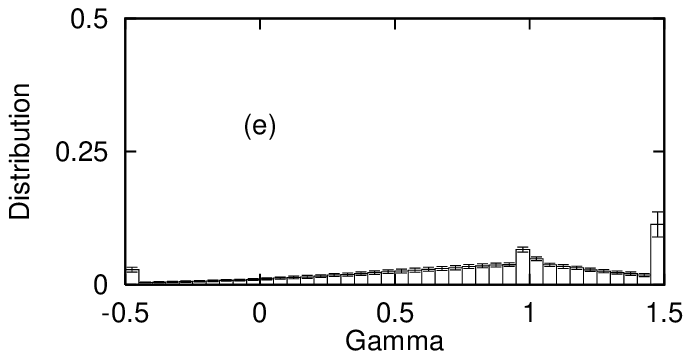}}}
\centerline{\hbox{\epsffile{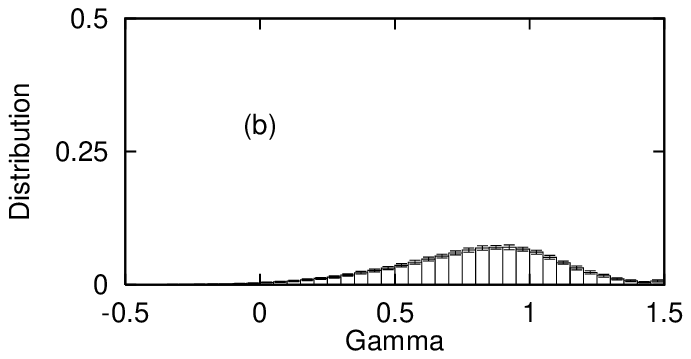} \epsffile{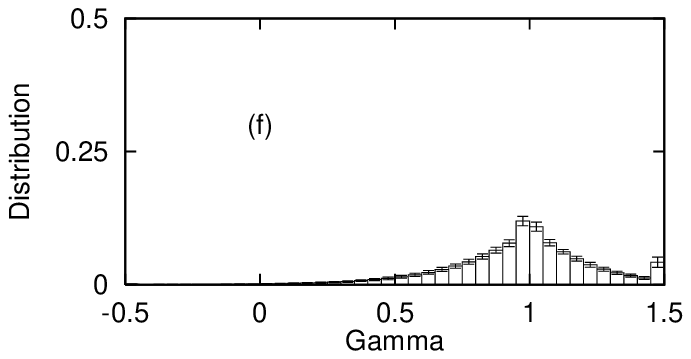}}}
\centerline{\hbox{\epsffile{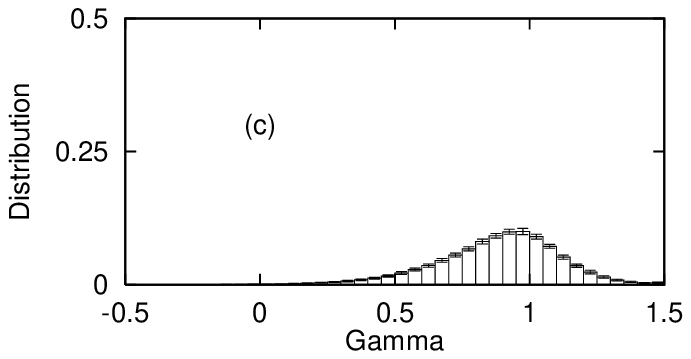} \epsffile{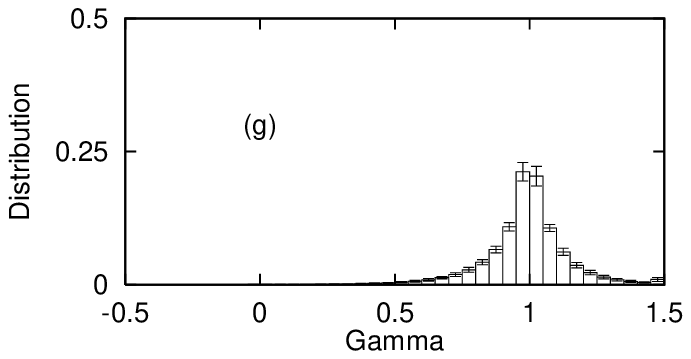}}}
\centerline{\hbox{\epsffile{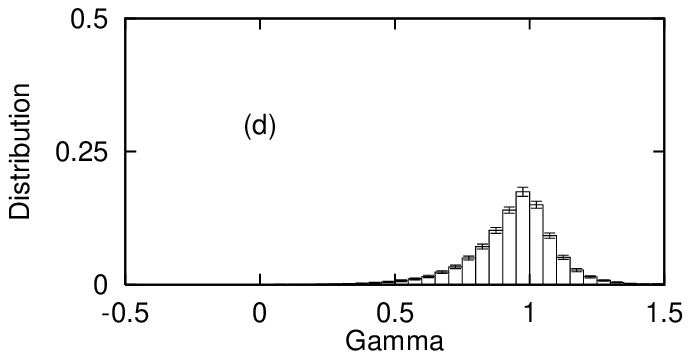} \epsffile{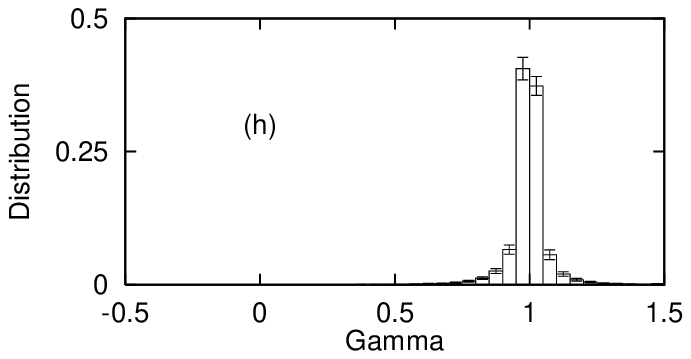}}}
\caption[3]{
The average number of stability coefficients $n_{\gamma}$ per interval of size
$0.05$, divided by the total number of the stability coefficients
$\gamma_i^{\mu,m}$, given by $NpL$, has been plotted for a neural network with dilution $0.6$, after
one or more learning cycles, for the non-local and local energy saving learning rules.
\protect\\
The calculations have been performed for a tabula rasa network,
$w_{ij}(t_0)=0$, of $N=128$ neurons with vanishing thresholds
($\theta_i=0$).
An average has been taken of $L=100$ sets of $p=32$ patterns. The
average activity is $a=0.2$.
\protect\\
Figures (a-d). 
The average number of stability coefficients after $1$, $5$, $10$ and
$20$ learning cycles in case of the local energy saving learning rule (\protect\ref{dw-local})--(\ref{eta}).
\protect\\
Figures (e-h). 
The average number of stability coefficients after $1$, $5$, $10$ and
$20$ learning cycles in case of the non-local energy saving learning rule (\protect\ref{delta-lagrange}).
\protect\label{fig3}}
\end{figure}

In order to judge the performance of the network in case of storage of
more patterns, we now picture the distribution of the $\gamma_i^{\mu, m}$ over the real axis.
Ideally, all $\gamma_i^{\mu,m}$ should be equal to $\kappa=1$.
In figure~\ref{fig3} the distribution has been plotted for both the
non-local and local energy saving learning rule.
As one observes from figure~\ref{fig3}, some of the $\gamma$'s have
values smaller than one (and even negative) whereas others have
values larger than one.
This is due to the fact that storing in set $m$ a pattern $\XI^{\nu}$,
the $\gamma_i^{\mu,m}$'s of the other patterns $\XI^{\mu}$ ($\mu \neq
\nu$) are not taken into account in the learning step and as a
consequence can be enlarged or reduced in value.
We have chosen to put the number of $\gamma$'s with values outside the
plotted interval in the very first and the very last interval: see,
e.g., figure~\ref{fig3}e. 

The general conclusion is that the local energy saving learning rule, although in
principle approximative, is an excellent competitor of the non-local one.
After five learning cycles already the number of negative
$\gamma_i^{\mu,m}$ is negligible [see figures~\ref{fig3}b and \ref{fig3}f], and the
distribution of the $\gamma_i^{\mu,m}$'s are comparable.

We finally make some observations regarding other learning rules.
In view of (\ref{gamma-the-one}), the symmetric learning rule (\ref{sym})
yields the same values of the $\gamma$'s as in case of our asymmetric learning rule
(\ref{the-one}).
Hence, in particular, the whole analysis of this section holds true
for the symmetric learning rule as well.
In other words, although the changes $\Delta w_{ij}$ in the weights
$w_{ij}$ as given by the symmetric learning rule (\ref{sym}) are, of
course, different from those given by our asymmetric learning rule
(\ref{the-one}), the convergence properties --- studied here via the
$\gamma$'s --- are exactly the same for the symmetric learning rule
(\ref{sym}) and our asymmetric learning rule (\ref{the-one}).
The `wrong' asymmetric learning rule (\ref{asym}) does not work at
all, as has been explained at the end of section~\ref{hebb}.

\section{Summary}
\label{summary}
We have shown that two different arguments, a biological one
(section~\ref{hebb}) and a physical one (section~\ref{lagrange}) lead
to a Hebb rule of the same asymmetric form: compare
eqs. (\ref{the-one})--(\ref{other}) at the one hand and
eq. (\ref{delta-lagrange}) at the other hand.
A learning rule of this form is never, or at least not often, used in the physical
literature, which, in general, is less concerned with an accurate
modeling of a biological network.

The biological argument was largely based on the improbability of a
change of connections if the pre-synaptic neuron was inactive.
The physical argument was based on the expression (\ref{de}) for the energy
change, not on any ad-hoc cost-function like (\ref{energy}) as has been done
so far in the literature.
The local version of the energy saving Hebb rule
(\ref{delta-lagrange}), given by eqs. (\ref{dw-local})--(\ref{eta}),
may be relevant for biological systems.
It has been tested numerically in section~\ref{numerical}, and turns
out to yield storage of patterns in a satisfactory way: see in particular
figure~\ref{fig3}.

\ack{The authors are indebted to Bob van Dijk, Hugo Keizer, Hans Capel, Bernard
  Nienhuis, Wytse Wadman, Henk Spekreijse and the referees, all of
  which contributed in some way to this article in its present form.} 

\appendix

\section{Maximal efficient learning}
\label{efficient}

We shall here merely verify the maximal efficient learning rule, not
derive the rule, since the derivation closely parallels the one of
Linkevich \cite{linkevich}.
In view of the special constraints mentioned directly above eq. (\ref{dw-ideal}), eq. (\ref{linear-fp}) reduces to
\be
\label{linear-link}
\sum_{j \in V_i} w_{ij}(t)\xi_j^{\mu} = 2 \kappa \xi_i^{\mu} \, , \quad 
(\mu=1,\ldots,p) \, .
\ee
Similarly, the solution (\ref{w-opper}) of (\ref{linear-fp}) reduces to
\be
\label{w-opper-link}
w_{ij}(t) = \left\{ \begin{array}{ll} {\displaystyle N^{-1}
  \sum_{\mu,\nu=1}^p 2 \kappa \xi_i^{\mu} (C_i^{-1})^{\mu \nu}
  \xi_j^{\nu} } \, , & (j \in V_i) \\ 0 \, , & (j \in V_i^c) \end{array} \right.
\ee

In order to store a new pattern $\XI^{p+1}$, the new weights
  $w_{ij}(t')$ have to obey the equations  
\be
\label{linear-link-new}
\sum_{j \in V_i} w_{ij}(t')\xi_j^{\mu} = 2 \kappa \xi_i^{\mu} \, , \quad 
(\mu=1,\ldots,p+1) \, .
\ee
The weights $w_{ij}(t')$ are related to the weights $w_{ij}(t)$ by
\be
\label{w-new}
w_{ij}(t')= \left\{ \begin{array}{ll} w_{ij}(t) +\Delta w_{ij}(t) \, ,
& (j \in V_i) \\ w_{ij}(t) \, , & (j \in V_i^c) \, , \end{array} \right.
\ee
where the $w_{ij}(t)$ are the connections after storage of the patterns
$\XI^1,\ldots, \XI^p$ as given by equation (\ref{w-opper-link}) and
the $\Delta w_{ij}(t)$ are given by (\ref{dw-ideal}).

Inserting (\ref{w-new}) with (\ref{w-opper-link}) and (\ref{dw-ideal}) into
the left-hand side of (\ref{linear-link-new}) yields
\be
\sum_{j \in V_i} w_{ij}(t')\xi_j^{\mu} = \left\{ \begin{array}{ll}
{\displaystyle 2\kappa \xi_i^{\mu} + \sum_{j \in V_i} \Delta w_{ij}(t)\xi_j^{\mu}} \, ,
& (\mu=1,\ldots,p) \\  {\displaystyle 2 \kappa \xi_i^{\mu}} \, , & (\mu=p+1) \end{array} \right.
\ee
The right-hand side of these equations is equal to that of
(\ref{linear-link-new}) if
\be
\label{dw=0}
 \sum_{j \in V_i} \Delta w_{ij}(t)\xi_j^{\mu}=0 \, , \quad 
 (\mu=1,\ldots,p) \, .
\ee 
In order to show that (\ref{dw=0}) holds, we first decompose $\xi_j^{p+1}$ according to
\be
\label{decom}
\xi_j^{p+1}= \left\{ \begin{array}{ll} {\displaystyle \sum_{\mu=1}^p a^{\mu}
  \xi_j^{\mu} + \psi_j^{p+1}} \, , & (j \in V_i) \\ {\displaystyle \sum_{\mu=1}^p a_j^{\mu}
  \xi_j^{\mu} + \psi_j^{p+1}} \, , & (j \in V_i^c) \end{array} \right. \, ,
\ee 
where $a^{\mu}$, $a_j^{\mu}$ and $\psi_j^{p+1}$ have been taken such that
\footnote[7]{In case all connections may change in time, the index sets $V_i$ are
all equal to the set of all indices.
Then the equations (\ref{decom}) with $j \in V_i^c$ disappear and
(\ref{orth-cond}) amounts to the condition that the vector
$\PSI^{p+1}$ is orthogonal to the vectors $\XI^{\mu}$
($\mu=1,\ldots,p$).
Hence, in this particular case there are $p+N$ restrictions
(\ref{decom}) and (\ref{orth-cond}) for $p+N$ variables $a^{\mu}$ and
$\psi_j$.}
\be
\label{orth-cond}
\sum_{j \in V_i} \xi_j^{\mu} \psi_j^{p+1}=0 \, , \quad (\mu=1,\ldots,p)
\, .
\ee
Using (\ref{w-opper-link}) and (\ref{orth-cond}) one may prove the auxiliary relation
\be
\label{aux-cond}
\sum_{k \in V_j} w_{jk}(t) \psi_k^{p+1}=0 \, .
\ee   
The proof of (\ref{dw=0}) is now straightforward.
First, substitution of (\ref{dw-ideal}) into (\ref{dw=0}) yields
\be
\label{dw-prop}
\sum_{j \in V_i} \Delta w_{ij}(t)\xi_j^{\mu} \propto  \sum_{j \in V_i}
[ 2 \kappa \xi_j^{p+1} -\sum_{k \in V_j} w_{jk}(t) \xi_k^{p+1} ] \xi_j^{\mu} \, , \quad  (\mu=1,\ldots,p) \, .
\ee 
Then, substituting the decomposition (\ref{decom}) in (\ref{dw-prop}), and
using (\ref{linear-link}), (\ref{orth-cond}) and (\ref{aux-cond}) we
see that this expression vanishes, which proves (\ref{dw=0}).
Hence, the left-hand side of (\ref{linear-link-new}) equals the
right-hand side of (\ref{linear-link-new}) for a learning rule given
by (\ref{dw-ideal}).

\section{Modified method of the pseudo-inverse}
\label{pseudo-inverse}

Consider the $p$ sets of $N$ linear equations
\be
\label{lin-eq}
\sum_{j=1}^N w_{ij} x_j^{\mu} =a_i^{\mu} \, , \quad (i=1,\ldots,N;\mu=1,\ldots,p)
\, , 
\ee
where $x_j^{\mu}$ and $a_i^{\mu}$ are known constants
($j=1,\ldots,N;\mu=1,\ldots,p$).
The $N^2$ unknowns $w_{ij}$ are not determined as long as $p<N$.
Let $V_i$ be the subset of indices $j$ with the property that $w_{ij}$
is a solution of the set of equations (\ref{lin-eq}), and let the
complement of the set $V_i$ with respect to the
total set of indices ($1,\ldots,N$), denoted by $V_i^c$, contain the indices $j$
with the property that the $w_{ij}$ have the pre-described constant values
$b_{ij}$, i.e.,
\be
\label{extra}
w_{ij}=b_{ij} \, , \quad (j \in V_i^c) \, .
\ee
chosen in such a way that the system of equations (\ref{lin-eq}) does not
become incompatible.
If the set $V_i^c$ is empty, a solution of (\ref{lin-eq}) can be
obtained via the Moore-Penrose pseudo-inverse matrix \cite{personnaz,kohonen}.
We want to obtain a solution for $w_{ij}$ of
(\ref{lin-eq})--(\ref{extra}) in case $V_i^c$ is not empty, and the
pseudo-inverse matrix can not be used directly.
To that end, we construct a new set of equations, closely related to
(\ref{lin-eq})--(\ref{extra}), which can be solved via the
pseudo-inverse.
We will refer to this construction as the \emph{modified method of
  the pseudo-inverse}.

We first define a new set of variables $\tilde{w}_{ij}$ according to
\be
\label{w-tilde}
\tilde{w}_{ij}=w_{ij}-b_{ij} \, \quad (i,j=1,2,\ldots,N) \, ,
\ee
where $b_{ij}$ are arbitrary in case $j \in V_i$.
We then have
\be
\label{tilde}
\tilde{w}_{ij}= \left\{ \begin{array}{ll} w_{ij}-b_{ij} \, , & (j \in
V_i) \\ 0 \, , & (j \in V_i^c) \end{array} \right. 
\ee
The under-determined set of $pN$ linear equations
(\ref{lin-eq})--(\ref{extra}) can now be rewritten
\be
\label{lin-eq-tilde}
\sum_{j \in V_i} \tilde{w}_{ij} x_j^{\mu} =\tilde{a}_i^{\mu} \, , \quad (\mu=1,\ldots,p)
\, ,
\ee
where
\be
\label{tilde-a}
\tilde{a}_i^{\mu}=a_i^{\mu}-\sum_{j=1}^N b_{ij} x_j^{\mu} \, .
\ee
Note that (\ref{lin-eq-tilde}) cannot be solved with the help of the
pseudo-inverse, since the summation is only with respect to a
restricted set of indices $j \in V_i$.
We therefore consider a new set of $pN$ linear equations, namely
\be
\label{new-lin-eq}
\sum_{j=1}^N v_{ij} y_j^{\mu} =\tilde{a}_i^{\mu} \, , \quad (\mu=1,\ldots,p)
\, .
\ee
The relation of (\ref{new-lin-eq}) to (\ref{lin-eq-tilde}) can be made
clear by taking
\be
\label{y}
y_j^{\mu}= \left\{ \begin{array}{ll} x_j^{\mu} \, , & (j \in
V_i) \\ 0 \, , & (j \in V_i^c) \, , \end{array} \right. 
\ee
since then the set of equations (\ref{new-lin-eq}) for the $N^2$ unknowns
$v_{ij}$ ($i,j=1,2,\ldots,N$) becomes identical to the set of
equations (\ref{lin-eq-tilde}) for the unknown $\tilde{w}_{ij}$
($i=1,\ldots,N;j \in V_i$).
The equation (\ref{new-lin-eq}) can be solved with the help of the
pseudo-inverse.
The solution reads
\be
v_{ij}=\sum_{\mu, \nu=1}^p \tilde{a}_i^{\mu} (C^{-1})^{\mu \nu} y_j^{\nu} \, , \quad
(i,j=1,2.\ldots,N) \, ,
\ee
where $C^{\mu \nu}$ is the usual correlation matrix \cite{hemmen-kuhn}
\be
C^{\mu \nu}=\sum_{k=1}^N y_k^{\mu} y_k^{\nu} \, .
\ee
If we use (\ref{y}), the matrix $C^{\mu \nu}$ becomes what we have
called the `reduced correlation matrix', given by
\be
C_i^{\mu \nu}=\sum_{k \in V_i} x_k^{\mu} x_k^{\nu} \, .
\ee
The modified correlation matrix takes into account the modifications
in the usual correlation matrix due to the particular network
architecture as dictated by the index set $V_i$.
The solutions $v_{ij}$ become, using (\ref{y}),
\be
\label{sol-v}
v_{ij}= \left\{ \begin{array}{ll} \sum_{\mu,\nu=1}^p \tilde{a}_i^{\mu} (C_i^{-1})^{\mu \nu} x_j^{\nu} \, , & (j \in
V_i) \\ 0 \, , & (j \in V_i^c) \, , \end{array} \right. 
\ee
Hence, the solution (\ref{sol-v}) turns out to be compatible with (\ref{tilde})
for $j \in V_i^c$.
Putting now 
\be
\tilde{w}_{ij}=v_{ij} \, , \quad (i,j=1,2,\ldots,N) \, ,
\ee
we have obtained a solution for (\ref{lin-eq-tilde}), as follows by
comparing (\ref{new-lin-eq}) and (\ref{lin-eq-tilde}).
In this way we find, transforming back from $\tilde{w}_{ij}$ to
$w_{ij}$ with the help of (\ref{w-tilde}), and substituting (\ref{tilde-a}), the
final result for the solution of the under-determined set of equations
(\ref{lin-eq})--(\ref{extra})~:
\be
\label{sol-w}
w_{ij}= \left\{ \begin{array}{ll} {\displaystyle b_{ij}+ \sum_{\mu,\nu=1}^p [a_i^{\mu}
-\sum_{j=1}^N b_{ij}x_j^{\mu} ] (C_i^{-1})^{\mu \nu} x_j^{\nu} } \, , & (j \in
V_i) \\ b_{ij} \, , & (j \in V_i^c) \, , \end{array} \right. 
\ee
We recall that the $b_{ij}$ are arbitrary for $j \in V_i$, and
prescribed for $j \in V_i^c$.
Notice that the solution (\ref{sol-w}) is not unique because of the
arbitrary constants $b_{ij}$ ($j \in V_i$).

We want to solve (\ref{linear-fp}) for a network with changing
connections $w_{ij}$ if $ j \in V_i$ and non-changing connections if $
j \in V_i^c$.
Applying (\ref{sol-w}) with
\bea
x_i^{\mu} &=& \xi_i^{\mu} \nn\\
b_{ij} &=& w_{ij}(t_0) \, , \quad (i,j=1,2,\ldots,N) \nn\\
a_i^{\mu} &=& \kappa ( 2\xi_i^{\mu}-1) + \theta_i
\eea
we obtain at once (\ref{w-opper}).
We thus arrive at the observation that the energy saving solution
(\ref{w-opper}) coincides with the solution (\ref{sol-w}), obtained
with the help of the modified method of the pseudo-inverse.

\end{document}